\documentclass[pre,
twocolumn
]{revtex4-1}
\usepackage{graphicx}
\usepackage{amsmath, amssymb, amsthm}
\usepackage{epsfig, color}
\usepackage{systeme}
\usepackage{multirow, comment, enumerate}
\usepackage{hyperref}
\usepackage{enumitem}
\usepackage{esint}
\usepackage{lipsum}
\usepackage{bm}

\definecolor{myblue}{RGB}{0,50,200}
\hypersetup{
	colorlinks,
	citecolor=myblue,
	linkcolor=myblue,
	urlcolor=myblue
}

\allowdisplaybreaks

\newcommand{\mca}{\mathcal}
\newcommand{\mbb}{\mathbb}
\newcommand{\mrm}{\mathrm}
\newcommand{\avg}[1]{\langle #1\rangle}

\newcommand{\avgl}[1]{\left\langle #1\right\rangle}
\newcommand{\set}[1]{\lbrace #1\rbrace}
\newcommand{\bra}[1]{\left( #1 \right)}
\newcommand{\bras}[1]{\left[ #1 \right]}
\newcommand{\brab}[1]{\left\{ #1 \right\}}

\newcommand{\norm}[1]{\Vert #1 \Vert}
\newcommand{\bX}{\bm{\mathrm{X}}}
\newcommand{\bXf}[1]{\bm{\mathrm{X}}_{[#1]}}
\newcommand{\bXd}[1]{\bm{\mathrm{X}}^\dagger_{[#1]}}
\newcommand{\pp}{\partial}
\newcommand{\intf}{\int_{-\infty}^{\infty}}
\newcommand{\inth}{\int_{0}^{\infty}}
\newcommand{\dz}{\frac{d}{dz}}
\newcommand{\dzt}{\frac{d^2}{dz^2}}

\begin{document}
\title{Uncertainty relations for time-delayed Langevin systems}
\author{Tan Van Vu}
\email{tan@biom.t.u-tokyo.ac.jp}
\affiliation{Department of Information and Communication Engineering, Graduate School of Information Science and Technology, The University of Tokyo, Tokyo 113-8656, Japan}

\author{Yoshihiko Hasegawa}
\email{hasegawa@biom.t.u-tokyo.ac.jp}
\affiliation{Department of Information and Communication Engineering, Graduate School of Information Science and Technology, The University of Tokyo, Tokyo 113-8656, Japan}

\date{\today}

\begin{abstract}
The thermodynamic uncertainty relation, which establishes a universal trade-off between nonequilibrium current fluctuations and dissipation, has been found for various Markovian systems.
However, this relation has not been revealed for non-Markovian systems; therefore, we investigate the thermodynamic uncertainty relation for time-delayed Langevin systems.
We prove that the fluctuation of arbitrary dynamical observables is constrained by the Kullback--Leibler divergence between the distributions of the forward path and its reversed counterpart.
Specifically, for observables that are antisymmetric under time reversal, the fluctuation is bounded from below by a function of a quantity that can be identified as a generalization of the total entropy production in Markovian systems.
We also provide a lower bound for arbitrary observables that are odd under position reversal.
The term in this bound reflects the extent to which the position symmetry has been broken in the system and can be positive even in equilibrium.
Our results hold for finite observation times and a large class of time-delayed systems because detailed underlying dynamics are not required for the derivation.
We numerically verify the derived uncertainty relations using two single time-delay systems and one distributed time-delay system.
\end{abstract}

\maketitle

\section{Introduction}
In the last two decades, substantial progress has been made in stochastic thermodynamics relevant to describing small systems that fluctuate and are far from thermal equilibrium \cite{Sekimoto.1998.PTPS,Denis.2002.AP,Seifert.2005.PRL,Seifert.2012.RPP,Yannick.2015.PA,Jaco.2016.EPL,Gong.2016.PRL,Goldt.2017.PRL}.
The first and second laws of thermodynamics have been generalized for individual trajectory levels, and fluctuation theorems \cite{Gallavotti.1995.PRL,Jarzynski.1997.PRL,Seifert.2012.RPP} that express universal properties of the probability distributions of thermodynamic quantities such as work, heat, and entropy production have been derived.
This framework has been used to investigate various systems such as optical and colloidal particle systems and biochemical reaction networks \cite{Seifert.2012.RPP}.

In recent years, the thermodynamic uncertainty relation (TUR), which states that smaller current fluctuation cannot be attained without higher thermodynamic cost, has been found in various Markovian dynamical processes \cite{Barato.2015.PRL,Gingrich.2016.PRL,Pietzonka.2016.PRE,Polettini.2016.PRE,Horowitz.2017.PRE,Andreas.2018.JSM}.
The TUR was first proved for large-time limit using the large deviation theory \cite{Gingrich.2016.PRL}; later, it was found to be valid even for finite observation times \cite{Horowitz.2017.PRE}.
The general form of the TUR is represented by the following inequality:
\begin{equation}
\frac{\mrm{Var}[\jmath]}{\avg{\jmath}^2}\ge\frac{2k_{\mrm{B}}}{\Sigma},\label{eq:conventional.TUR}
\end{equation}
where $k_{\rm B}$ is Boltzmann's constant, $\avg{\jmath}$ and $\mrm{Var}[\jmath]$ are the mean and variance of the current, respectively, and $\Sigma$ is the average of the total entropy production.
Analogous precision-cost trade-off relations have been reported in the literature \cite{Mehta.2012.PNAS,Hasegawa.2018.PRE}.
Various forms of the TUR have been proposed and studied intensively in many other contexts \cite{Barato.2015.JPCB,Falasco.2016.PRE,Patrick.2016.JSM,Rotskoff.2017.PRE,Garrahan.2017.PRE,Gingrich.2017.PRL,Hyeon.2017.PRE,Proesmans.2017.EPL,Chiuchiu.2018.PRE,Brandner.2018.PRL,Hwang.2018.JPCL,Barato.2018.NJP,Barato.2018.arxiv,Hasegawa.2019.PRE,Koyuk.2019.JPA,Vu.2019.arxiv}.
Hereafter, the term ``bound'' refers the lower bound on the relative fluctuation of currents.

To date, the TUR has only been investigated in Markovian systems.
However, the time delay that causes non-Markovian dynamical behavior inevitably exists in many real-world stochastic processes such as gene regulation \cite{Bratsun.2005.PNAS,Mather.2009.PRL}, biochemical reaction networks \cite{Novak.2008.NAT}, and control systems involving a feedback protocol \cite{Kim.1999.PRL,Masoller.2003.PRL,Lichtner.2012.PRE}.
It is well-known that time delay can completely alter system dynamics, e.g., delay-induced oscillations \cite{Bratsun.2005.PNAS}.
Recently, Ref.~\cite{Loos.2019.SR} has shown that even a small delay time leads to finite heat flow in the system.
Despite the importance of delay in many classical and quantum systems, thermodynamic analysis of such systems remains challenging \cite{Rosinberg.2015.PRE,Rosinberg.2017.PRE}.

In this paper, we study the TUR for general dynamical observables that are antisymmetric under conjugate operations such as time or position reversal.
First, we define a trajectory-dependent quantity $\sigma$ [cf.~Eq.~\eqref{eq:sigma.definition}], whose average is the Kullback--Leibler (KL) divergence between the distributions of the forward path and its conjugate counterpart.
In the absence of time delay and under time reversal, $\sigma$ is identified as the trajectory-dependent total entropy production in Markovian systems.
Starting from the point that the joint probability distribution of $\sigma$ and the observable obeys the strong detailed fluctuation theorem, we prove that the relative fluctuation of the observable is lower bounded by $2/(e^{\avg{\sigma}}-1)$.
This implies that the time irreversibility in the system constrains the fluctuation of observables that are odd under time reversal.
For observables that are antisymmetric under position reversal, the bound reflects the degree of position-symmetry breaking in the system.
The derived bound holds for arbitrary observation times and for a large class of time-delayed systems such as continuous- or discrete-time systems with multiple or distributed delays.
We numerically verify the validity of the derived inequality in three systems wherein $\avg{\sigma}$ can be analytically obtained.

\section{Model}
To clearly illustrate the results, we consider here a single time-delayed system with dynamical variables $\bm{x}(t)=[x_1(t),\dots,x_N(t)]^\top$, as described by the following set of coupled Langevin equations:
\begin{equation}
\dot{\bm{x}} = \bm{F}(\bm{x},\bm{x}_\tau)+\bm{\xi},\label{eq:single.timedelay.Langevin}
\end{equation}
where $\bm{x}_\tau\equiv\bm{x}(t-\tau)$, $\bm{F}(\bm{x},\bm{x}_\tau)\in\mbb{R}^N$ is a drift force, $\bm{\xi}(t)=[\xi_1(t),\dots,\xi_N(t)]^\top$ is zero-mean white Gaussian noise with covariance $\avg{\xi_i(t)\xi_j(t')}=2D_i\delta_{ij}\delta(t-t')$, and $\tau\ge 0$ denotes the delay time in the system.
Here, $D_i$'s denote the noise intensity.
Equation~\eqref{eq:single.timedelay.Langevin} is interpreted as Ito stochastic integration.
Throughout this paper, Boltzmann's constant is set to $k_{\rm B}=1$.
Let $P(\bm{x},t)$ be the probability distribution function for the system to be in state $\bm{x}$ at time $t$.
Then, the corresponding Fokker-Planck equation (FPE) is expressed as \cite{Frank.2003.PRE,Frank.2005.PRE}
\begin{equation}
\partial_t P(\bm{x},t)=-\sum_{i=1}^N\partial_{x_i}J_i(\bm{x},t),\label{eq:general.FPE}
\end{equation}
where
\begin{equation}
\begin{aligned}
J_i(\bm{x},t) &=\int d\bm{y}\,F_i(\bm{x},\bm{y})P(\bm{y}, t-\tau;\bm{x},t)-D\partial_{x_i}P(\bm{x},t)\\
&=\overline{F}_i(\bm{x}) P(\bm{x},t) - D_i\partial_{x_i}P(\bm{x},t)
\end{aligned}
\end{equation}
is the probability current. Here,
\begin{equation}
\overline{F}_i(\bm{x}) = \int d\bm{y}\,F_i(\bm{x},\bm{y})P(\bm{y},t-\tau |\bm{x},t)
\end{equation}
is an effective force obtained by taking the delay-averaged integration of the variable $\bm{y}$ and $P(\bm{y},t-\tau;\bm{x},t)$ is a joint probability density for a system that takes value $\bm{x}$ at time $t$ and $\bm{y}$ at time $t-\tau$.
Generally, solving $P(\bm{y},t-\tau;\bm{x},t)$ results in an infinite hierarchy of equations, where $n$-time probability distribution depends on the $(n+1)$-time one.
Therefore, it is difficult to analytically obtain the effective force $\overline{F}_i(\bm{x})$, except in linear systems.

We define $\bXf{s,e}\equiv\set{\bm{x}(t)}_{t=s}^{t=e}$ as a trajectory that begins at $t=s$ and ends at $t=e$.
Let $\mca{P}(\bXf{s,e})$ be the probability of observing the trajectory $\bXf{s,e}$.
For each trajectory $\bXf{s,e}$, we consider a conjugate trajectory $\bXd{s,e}$ defined by $\bXd{s,e}\equiv\set{\bm{x}^\dagger(t)}_{t=s}^{t=e}$.
Assuming that we observe the system during a time interval $[0,T]$, we then define a trajectory-dependent quantity $\sigma(\bXf{0,T})$, which is the ratio of the probabilities of observing the forward path and its conjugate counterpart, as follows:
\begin{equation}
\sigma\equiv\ln\frac{\mca{P}(\bXf{0,T})}{\mca{P}(\bXd{0,T})}.\label{eq:sigma.definition}
\end{equation}
For the sake of simplicity, we use the notation $\bX$, omitting the time interval, to indicate $\bXf{0,T}$.
If the conjugate operation satisfies the property $(\bX^{\dagger})^{\dagger}=\bX$, then $\sigma$ is odd under it, i.e., $\sigma(\bX^\dagger)=-\sigma(\bX)$.
Hereafter, we consider conjugate operations that satisfy this property.
Introducing the probability distribution $P(\sigma)=\int\mca{D}\bX\,\delta(\sigma-\sigma(\bX))\mca{P}(\bX)$, we show that $P(\sigma)$ satisfies the fluctuation theorem, i.e.,
\begin{equation}
\frac{P(\sigma)}{P(-\sigma)}=e^{\sigma}.\label{eq:strong.DFT}
\end{equation}
Equation~\eqref{eq:strong.DFT} can be derived as follows:
\begin{align}
P(\sigma)&=\int\mca{D}\bX\,\delta(\sigma-\sigma(\bX))\mca{P}(\bX)\nonumber\\
&=\int\mca{D}\bX\,\delta(\sigma-\sigma(\bX))e^{\sigma(\bX)}\mca{P}(\bX^\dagger)\nonumber\\
&=e^{\sigma}\int\mca{D}\bX\,\delta(\sigma-\sigma(\bX))\mca{P}(\bX^\dagger)\nonumber\\
&=e^{\sigma}\int\mca{D}\bX^\dagger\,\delta(\sigma+\sigma(\bX^\dagger))\mca{P}(\bX^\dagger)\nonumber\\
&=e^{\sigma}P(-\sigma).
\end{align}
Equation \eqref{eq:strong.DFT} implies that $\sigma$ satisfies the integral fluctuation theorem, i.e., $\avg{e^{-\sigma}}=1$.
By applying Jensen's inequality $\avg{e^{-\sigma}}\ge e^{-\avg{\sigma}}$, we have $\avg{\sigma}\ge 0$.
The average value of $\sigma$ can also be interpreted as the KL divergence between the distributions $\mca{P}$ and $\mca{P}^\dagger$
\begin{equation}\label{eq:avg.sigma.KL}
\avg{\sigma}=\mca{D}_{\rm KL}[\mca{P}||\mca{P}^\dagger]=\int\mca{D}\bX\,\mca{P}(\bX)\ln\frac{\mca{P}(\bX)}{\mca{P}^\dagger(\bX)},
\end{equation}
where $\mca{P}^\dagger(\bX)\equiv\mca{P}(\bX^\dagger)$.
From Eq.~\eqref{eq:avg.sigma.KL}, $\avg{\sigma}$ becomes zero only when $\mca{P}(\bX)=\mca{P}(\bX^\dagger)$ for all trajectories $\bX$.
\begin{figure}[t]
	\centering
	\includegraphics[width=8.5cm]{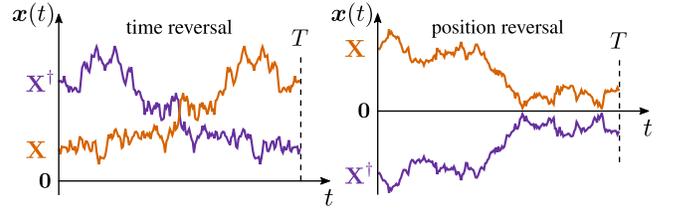}
	\protect\caption{Illustration of the conjugate operations. For simplicity, we assume here that the system involves only even variables. For the trajectory $\bX\equiv\set{x(t)}_{t=0}^{t=T}$, its reversed counterpart is $\bX^\dagger\equiv\set{x^\dagger(t)}_{t=0}^{t=T}$, where $x^\dagger(t)$ is equal to $x(T-t)$ (or $-x(t)$) under time reversal (or position reversal). Here, $T$ denotes the observation time.}\label{fig:conjugate.illustration}
\end{figure}

Let us discuss the conjugate operations that will be used here.
The most conventional one is time reversal, i.e., $\bm{x}^\dagger(t)=\bm{\epsilon}\bm{x}(T-t)$.
Here, $\epsilon_i=\pm 1$ for even and odd variables $x_i$, respectively.
For systems where both even and odd variables exist, a reversed trajectory $\bX^\dagger$ can be generated under forward dynamics.
Therefore, $\sigma$ is mathematically well defined.
In this case, $\avg{\sigma}$ is a measure of the time-reversal symmetry breaking in the system.
For steady-state systems involving only even variables, $\sigma$ can be decomposed as
\begin{equation}\label{eq:sigma.decomp.general}
\sigma=-\ln\frac{P^{\rm ss}(\bm{x}(T))}{P^{\rm ss}(\bm{x}(0))}+\ln\frac{\mca{P}(\bX|\bm{x}(0))}{\mca{P}(\bX^{\dagger}|\bm{x}^{\dagger}(0))},
\end{equation}
where $P^{\rm ss}(\cdot)$ is the steady-state distribution and $\mca{P}(\cdot|\cdot)$ is the conditional path probability.
When the time delay vanishes, $\sigma$ is identified as the total entropy production along a trajectory in Markovian systems \cite{Seifert.2012.RPP}; the first and second terms in the right-hand side of Eq.~\eqref{eq:sigma.decomp.general} correspond to the system and medium entropy production, respectively.
Under time reversal, $\avg{\sigma}$ can be considered a generalization of total entropy production for time-delayed systems \cite{GomezMarin.2008.PRE,Roldan.2012.PRE}.
It is worth noting that this generalization of entropy production is mathematical and that it is generally difficult to assess its relation to the thermodynamic notion of entropy production \cite{Diana.2014.JSM}, except in Markovian processes where an explicit connection was established \cite{Gaspard.2004.JSP,Seifert.2005.PRL}.
Another possible conjugate operation is position reversal, i.e., $\bm{x}^\dagger(t)=\bm{\kappa}-\bm{x}(t)$.
Here, $\bm{\kappa}\in\mbb{R}^{N}$ is a constant that can basically take an arbitrary value, except in systems involving $n^{\rm th}$-time-derivative variables, where $n\in\mbb{N}_{>0}$.
For these systems, $\bm{\kappa}$ must be carefully chosen to ensure that a reversed trajectory can be generated by forward dynamics.
In particular, $\bm{\kappa}$ must be set to $\kappa_i=0$ for all such variables $x_i$.
For example, if the system variables are the position and velocity of a particle, i.e., $\bm{x}(t)=[r(t),\dot{r}(t)]^\top$, where $r(t)$ is the particle's position, then the reversed trajectory $\set{\bm{x}^\dagger(t)}=\set{\kappa_1-r(t),\kappa_2-\dot{r}(t)}$ can be generated by the forward dynamics only if $\kappa_2=0$.
Under this conjugate operation, $\avg{\sigma}$ reflects the degree of position-symmetry breaking with respect to the position $\bm{\kappa}/2$ in the system.
In the remaining part of the paper, we consider the $\bm{\kappa}=0$ case.
To distinguish when each operation is employed, we use subscripts $t$ and $p$ to refer time reversal and position reversal, respectively.
The conjugate operations are illustrated in Fig.~\ref{fig:conjugate.illustration}.

Because $\avg{\sigma}$ is the KL divergence between forward and reversed trajectories and trajectory-based quantities were previously measured \cite{Trepagnier.2004.PNAS,Collin.2005.Nature,Schuler.2005.PRL,Tietz.2006.PRL,Andrieux.2007.PRL}, $\avg{\sigma}$ is in principle experimentally measurable.
As will be shown in the examples, $\avg{\sigma}$ can be analytically calculated for several classes of systems.
In what follows, we investigate a more detailed form of $\sigma$ with respect to conjugate operations for the system defined in Eq.~\eqref{eq:single.timedelay.Langevin}.
For $T>\tau$, the path probability can be rewritten
\begin{equation}\label{eq:path.probability.decomposition}
\begin{aligned}
\mca{P}(\bXf{0,T})&=\mca{P}(\bXf{\tau,T}|\bXf{0,\tau})\mca{P}(\bXf{0,\tau}),\\
\mca{P}(\bXd{0,T})&=\mca{P}(\bXd{\tau,T}|\bXd{0,\tau})\mca{P}(\bXd{0,\tau}),
\end{aligned}
\end{equation}
where $\mca{P}(\bXf{\tau,T}|\bXf{0,\tau})$ is the probability of observing $\bXf{\tau,T}$, conditioned on $\bXf{0,\tau}$.
We note that under time reversal, $\bXd{0,\tau}=\set{\bm{\epsilon}\bm{x}(T-t)}_{t=0}^{t=\tau}$.
The conditional probability can be calculated via the path integral as
\begin{equation}
\mca{P}(\bXf{\tau,T}|\bXf{0,\tau})=\mca{N}\exp\bra{-\sum_{i=1}^N\mca{S}_i(\bXf{0,T})},\label{eq:path.integral}
\end{equation}
where $\mca{S}_i(\bXf{0,T})$ is the stochastic action given by
\begin{equation}
\mca{S}_i(\bXf{0,T})=\int_{\tau}^{T}dt\bras{\frac{\bra{\dot{x}_i-F_i(\bm{x},\bm{x}_\tau)}^2}{4D_i}+\frac{\pp_{x_i}F_i(\bm{x},\bm{x}_\tau)}{2}},\label{eq:stochastic.action}
\end{equation}
and $\mca{N}$ is a positive term independent of the trajectory.
Equation~\eqref{eq:path.integral} can be obtained by discretizing the Langevin equation [cf. Eq.~\eqref{eq:single.timedelay.Langevin}] and evaluating the path probability via the occurrence probability of the noise trajectory \cite{Aron.2010.JSM}.
The cross term $\int dt\,F_i(\bm{x},\bm{x}_\tau)\dot{x}_i$ in Eq.~\eqref{eq:stochastic.action} should be interpreted as $\int dt\,F_i(\bm{x},\bm{x}_\tau)\circ\dot{x}_i$, where $\circ$ denotes the Stratonovich product.
Using Eq.~\eqref{eq:path.probability.decomposition}, the average of $\sigma$ can be decomposed as
\begin{equation}
\avg{\sigma}=\avgl{\ln\frac{\mca{P}(\bXf{\tau,T}|\bXf{0,\tau})}{\mca{P}(\bXd{\tau,T}|\bXd{0,\tau})}}+\avgl{\ln\frac{\mca{P}(\bXf{0,\tau})}{\mca{P}(\bXd{0,\tau})}}.\label{eq:sigma.decomposition}
\end{equation}
In the long-time limit, i.e., $T\to\infty$, the first term in the right-hand side of Eq.~\eqref{eq:sigma.decomposition} becomes dominant as the second term is only a boundary value.
Neglecting the contribution of this boundary term and plugging Eq.~\eqref{eq:path.integral} into Eq.~\eqref{eq:sigma.decomposition}, $\avg{\sigma_{t}}$ and $\avg{\sigma_{p}}$ can be approximated as
\begin{widetext}
\begin{equation}\label{eq:avg.sigma}
\begin{aligned}
\avg{\sigma_{t}}&\approx\frac{1}{2}\sum_{i=1}^N\avgl{\int_{0}^{T-\tau}dt\bras{\frac{\bra{\dot{x}_i+F_i(\bm{x},\bm{x}_{-\tau})}^2}{2D_i}+\pp_{x_i}F_i(\bm{x},\bm{x}_{-\tau})}-\int_{\tau}^{T}dt\bras{\frac{\bra{\dot{x}_i-F_i(\bm{x},\bm{x}_\tau)}^2}{2D_i}+\pp_{x_i}F_i(\bm{x},\bm{x}_\tau)}},\\
\avg{\sigma_{p}}&\approx\frac{1}{2}\sum_{i=1}^N\avgl{\int_{\tau}^{T}dt\,\bras{\bra{\frac{\dot{x}_i}{D_i}-\frac{F_i(\bm{x},\bm{x}_\tau)-F_i(-\bm{x},-\bm{x}_\tau)}{2D_i}-\pp_{x_i}}\circ\bra{F_i(\bm{x},\bm{x}_\tau)+F_i(-\bm{x},-\bm{x}_\tau)}}}.
\end{aligned}
\end{equation}
\end{widetext}
For general systems, it is difficult to obtain more detailed forms of $\avg{\sigma_{t}}$ and $\avg{\sigma_{p}}$ than those in Eq.~\eqref{eq:avg.sigma}, except in linear systems.
$\avg{\sigma_{t}}$ becomes zero when the system is in equilibrium because $\avg{\sigma_{t}}$ characterizes the time reversibility of the system.
Contrastingly, $\avg{\sigma_{p}}$ can be positive even in the equilibrium system so long as the symmetry with respect to position reversal is broken.

\section{Derivation of uncertainty relation}
In this section, we derive the TUR for an arbitrary dynamical observable $\jmath(\bX)$, which is antisymmetric under the conjugate operation, i.e., $\jmath(\bX^\dagger)=-\jmath(\bX)$.
This antisymmetric property can be satisfied, e.g., for generalized currents of the form $\jmath(\bX)=\int_0^Tdt\,\bm{\Lambda}(\bm{x})^\top\circ\dot{\bm{x}}$ under time reversal, or for observables $\jmath(\bX)=\int_0^Tdt\,\Gamma_{\rm o}(\bm{x})$ or $\jmath(\bX)=\int_0^Tdt\,\bm{\Gamma}_{\rm e}(\bm{x})^\top\circ\dot{\bm{x}}$ under position reversal.
Here, $\Gamma_{\rm o}(\bm{x})$ and $\bm{\Gamma}_{\rm e}(\bm{x})$ are arbitrary odd and even functions, respectively.

In Ref.~\cite{Hasegawa.2019.arxiv}, we derived a modified variant of the TUR using the fluctuation theorem for Markovian processes.
Regardless of the underlying dynamics, the bound holds for as long as the fluctuation theorem is valid.
Here, we apply the same technique and derive the TUR for time-delayed systems.
First, we show that the joint probability distribution of $\sigma$ and $\jmath$, $P(\sigma,\jmath)$, obeys the fluctuation theorem; this can be proved analogously as follows:
\begin{align}
P(\sigma,\jmath)&=\int\mca{D}\bX\,\delta(\sigma-\sigma(\bX))\delta(\jmath-\jmath(\bX))\mca{P}(\bX)\nonumber\\
&=\int\mca{D}\bX\,\delta(\sigma-\sigma(\bX))\delta(\jmath-\jmath(\bX))e^{\sigma(\bX)}\mca{P}(\bX^\dagger)\nonumber\\
&=e^{\sigma}\int\mca{D}\bX\,\delta(\sigma-\sigma(\bX))\delta(\jmath-\jmath(\bX))\mca{P}(\bX^\dagger)\nonumber\\
&=e^{\sigma}\int\mca{D}\bX^\dagger\,\delta(\sigma+\sigma(\bX^\dagger))\delta(\jmath+\jmath(\bX^\dagger))\mca{P}(\bX^\dagger)\nonumber\\
&=e^{\sigma}P(-\sigma,-\jmath)\label{eq:strong.joint.DFT}.
\end{align}
Inspired by Ref.~\cite{Merhav.2010.JSM}, where the statistical properties of entropy production were obtained from the strong detailed fluctuation theorem, we derive the TUR solely from Eq.~\eqref{eq:strong.joint.DFT}.
Based on the following relation:
\begin{align}
1&=\intf d\sigma\intf d\jmath\, P(\sigma,\jmath)\nonumber\\
&=\inth d\sigma\intf d\jmath\,(1+e^{-\sigma})P(\sigma,\jmath),
\end{align}
we introduce a probability distribution $Q(\sigma,\jmath)\equiv(1+e^{-\sigma})P(\sigma,\jmath)$, defined over $[0,\infty)\times(-\infty,\infty)$.
Using the distribution $Q(\sigma,\jmath)$, the moments of $\sigma$ and $\jmath$ can be expressed in an alternative way as follows:
\begin{equation}
\begin{aligned}
\avg{\sigma^{2k}}&=\avgl{\sigma^{2k}}_{Q},\\
\avg{\jmath^{2k}}&=\avgl{\jmath^{2k}}_{Q},\\
\avg{\sigma^{2k+1}}&=\avgl{\sigma^{2k+1}\tanh\bra{\frac{\sigma}{2}}}_{Q},\\
\avg{\jmath^{2k+1}}&=\avgl{\jmath^{2k+1}\tanh\bra{\frac{\sigma}{2}}}_{Q},
\end{aligned}
\end{equation}
where $\avg{\cdot\cdot}_{Q}$ denotes the expectation with respect to $Q(\sigma,\jmath)$.
Applying the Cauchy-Schwartz inequality to $\avg{\jmath}$, we obtain
\begin{equation}
\avg{\jmath}^2=\avgl{\jmath\tanh\bra{\frac{\sigma}{2}}}^2_{Q}\le\avg{\jmath^2}_{Q}\avgl{\tanh\bra{\frac{\sigma}{2}}^2}_{Q}.\label{eq:bound1}
\end{equation}
The last term in the right-hand side of Eq.~\eqref{eq:bound1} can be further upper bounded.
We find that
\begin{equation}
\avgl{\tanh\bra{\frac{\sigma}{2}}^2}_{Q}\le\tanh\bra{\frac{\avg{\sigma}}{2}}.\label{eq:bound2}
\end{equation}
Equation ~\eqref{eq:bound2} is obtained by first noticing that $\tanh\bra{\frac{\sigma}{2}}^2\le\tanh\bras{\frac{\sigma}{2}\tanh\bra{\frac{\sigma}{2}}}$ for all $\sigma\ge 0$.
Thereafter, by applying Jensen's inequality to the concave function $\tanh(x)$, we obtain
\begin{align}
\avgl{\tanh\bras{\frac{\sigma}{2}\tanh\bra{\frac{\sigma}{2}}}}_{Q}&\le\tanh\bra{\avgl{\frac{\sigma}{2}\tanh\bra{\frac{\sigma}{2}}}_{Q}}\nonumber\\
&=\tanh\bra{\frac{\avg{\sigma}}{2}}.
\end{align}
From Eqs.~\eqref{eq:bound1} and \eqref{eq:bound2}, we have
\begin{equation}
\avg{\jmath}^2\le\avg{\jmath^2}\tanh\bra{\frac{\avg{\sigma}}{2}}.\label{eq:bound3}
\end{equation}
By transforming Eq.~\eqref{eq:bound3}, we obtain the following TUR for the observable $\jmath$:
\begin{equation}
\frac{\mrm{Var}[\jmath]}{\avg{j}^2}=\frac{\avg{\jmath^2}-\avg{\jmath}^2}{\avg{\jmath}^2}\ge\frac{2}{e^{\avg{\sigma}}-1}.\label{eq:main.TUR}
\end{equation}
The inequality in Eq.~\eqref{eq:main.TUR} is the main result of the paper.
For observables that are antisymmetric under time (or position) reversal, the term $\avg{\sigma}$ in the bound should be replaced by $\avg{\sigma_{t}}$ (or $\avg{\sigma_{p}}$).

In the limit $\tau\to 0$, the system [cf.~Eq.~\eqref{eq:single.timedelay.Langevin}] becomes a continuous-time Markovian process, with the conventional TUR providing a lower bound on the current fluctuations as in Eq.~\eqref{eq:conventional.TUR}.
Since $e^{\avg{\sigma}}-1\ge\avg{\sigma}$, the derived bound is looser than the conventional bound.
Regarding this difference, there are two possible explanations.
Firstly, it is because there is no requirement on the details of the underlying dynamics of the system considered in the derivation.
It was proven that the conventional bound does not hold for discrete-time Markovian processes \cite{Shiraishi.2017.arxiv,Proesmans.2017.EPL}.
Contrastingly, the derived bound holds for both continuous- and discrete-time systems.
The lower bound in Eq.~\eqref{eq:main.TUR} is the same as that in Ref.~\cite{Proesmans.2017.EPL} in which the TUR was derived in the long-time limit for discrete-time Markovian processes.
Secondly, the derived bound also holds for non-current observables, and differs from the conventional bound that holds only for current-type observables defined by $\jmath(\bX)=\int_0^Tdt\,\bm{\Lambda}(\bm{x})^\top\circ\dot{\bm{x}}$.

\section{Examples}
In this section, we study the derived bound with the help of three systems.
The first two steady-state systems are embedded in a Markovian heat reservoir, whereas the third is in contact with a non-Markovian environment, i.e., a heat reservoir with memory effects.
Unlike the conventional TUR, which was derived for steady-state systems, our bound holds even for non-steady-state systems.
Therefore, in the last system, we focus on a non-steady state.
For steady-state systems, $P^{\rm ss}(\bm{x})$ and $\bm{J}^{\rm ss}(\bm{x})$ denote the probability distribution and the probability current, respectively.

\subsection{One-dimensional system}
We study a one-dimensional linear system whose drift term is given by
\begin{equation}
F(x,x_\tau)=-ax-bx_\tau+f,\label{eq:Langevin.exa1}
\end{equation}
where $a$, $b$, and $f$ are the given constants satisfying the conditions $a>b>0,~f>0$.
It is easy to see that $\avg{x}=\overline{f}$, where $\overline{f}=f/(a+b)$.
The system has a Gaussian steady-state distribution that exists for arbitrary delay time $\tau$ because the force is linear, .
We introduce a new stochastic variable $z$, defined as $z=x-\overline{f}$.
The FPE corresponding to $z$ reads as
\begin{equation}
\partial_tP(z,t)=-\pp_{z}[\overline{G}(z)P(z,t)]+D\pp_{z}^2P(z,t),\label{eq:FPE.exa1}
\end{equation}
where $\overline{G}(z)=\int dy\bra{-az-by}P(y,t-\tau|z,t)$.
At the steady state, the probability current vanishes, i.e., $J^{\rm ss}(z)=\overline{G}(z)P^{\rm ss}(z)-D\pp_zP^{\rm ss}(z)=0$.
Here, $P^{\rm ss}(z)$ denotes the steady-state distribution.
Let $\phi(t)=\avg{z(0)z(t)}$ be the time-correlation function of $z$; it was shown that $\phi(t)=A_{+}e^{-c|t|}+A_{-}e^{c|t|}$ for all $|t|\le\tau$ \cite{Kuchler.1992.SSR,Frank.2003.PRE}, where $c=\sqrt{a^2-b^2},~A_{\pm}=1/2\bras{\phi(0)\pm D/c}$, and
\begin{equation}
\phi(0)=\avg{z^2}=\frac{D}{c}\frac{c+b\sinh(c\tau)}{a+b\cosh(c\tau)}.
\end{equation}
First, we consider the TUR for observables that are antisymmetric under time reversal.
According to Eq.~\eqref{eq:main.TUR}, the following inequality should be satisfied:
\begin{equation}
\frac{\avg{\jmath}^2}{\mrm{Var}[\jmath]}\le\frac{e^{\avg{\sigma_{t}}}-1}{2}.\label{eq:time.TUR.exa1}
\end{equation}
Since evaluating $\avg{\sigma_{t}}$ for $T>\tau$ necessitates complicated calculations, we consider only the case of $T\le\tau$ in which the path probability $\mca{P}(\bXf{0,T})$ can be calculated analytically as \cite{Rosinberg.2015.PRE}
\begin{align}
&\mca{P}(\bXf{0,T})\propto\exp\bras{-\frac{1}{4D}\int_{0}^{T}dt\bra{\dot{x}+cx-c\overline{f}}^2}\times\\
&\exp\bra{-\frac{c}{2D}\frac{\bras{A_+e^{-cT}\bra{x(0)-\overline{f}}-A_{-}\bra{x(T)-\overline{f}}}^2}{A_+^2e^{-2cT}-A_{-}^2}}.\nonumber
\end{align}
It can be confirmed that $\mca{P}(\bX)=\mca{P}(\bX^\dagger)$; thus, $\avg{\sigma_{t}}=0$.
Consequently, Eq.~\eqref{eq:time.TUR.exa1} implies that an arbitrary observable that is antisymmetric under time reversal vanishes on average, i.e., $\avg{\jmath}=0$.
For the current-type observable defined by $\jmath(\bX)=\int_0^{T}dt\,\Lambda(x)\circ\dot{x}(t)$, where $\Lambda(x)$ is an arbitrary projection function, one can easily check that $\avg{\jmath}=T\int_{-\infty}^{\infty}dz\,\Lambda(z+\overline{f})J^{\rm ss}(z)=0$.
Generally, this can be proven as
\begin{equation}
\begin{aligned}
 \avg{\jmath}&=\int\mca{D}\bX\,\jmath(\bX)\mca{P}(\bX)\\
 &=\frac{1}{2}\bra{\int\mca{D}\bX\,\jmath(\bX)\mca{P}(\bX)-\int\mca{D}\bX^\dagger\,\jmath(\bX^\dagger)\mca{P}(\bX^\dagger)}=0.
\end{aligned}
\end{equation}

Next, let us consider the TUR for non-current observables that are antisymmetric under position reversal.
Specifically, we validate the TUR for the observable $\jmath(\bX)=\int_0^Tdt\,x$, representing the area under the trajectory.
The average of the observable is $\avg{\jmath}=T\avg{x}=T\overline{f}$.
For $T\le\tau$, using the path integral, $\sigma_{p}$ can be calculated as
\begin{equation}\label{eq:pos.sigma.exa1}
\begin{aligned}
\sigma_{p}=\ln\frac{\mca{P}(\bXf{0,T})}{\mca{P}(\bXd{0,T})}&=\frac{c\overline{f}}{D}\int_{0}^{T}dt\,\bra{\dot{x}+cx}\\
&+\frac{2c\overline{f}}{D}\frac{A_+e^{-cT}x(0)-A_{-}x(T)}{A_+e^{-cT}+A_{-}}.
\end{aligned}
\end{equation}
Because the system is in the steady state, we obtain
\begin{equation}
\avg{\sigma_{p}}=\bra{cT+2\frac{A_+e^{-cT}-A_{-}}{A_+e^{-cT}+A_{-}}}\frac{c\overline{f}^2}{D}.\label{eq:avg.pos.sigma.smallT}
\end{equation}
The variance of the observable can also be obtained analytically as follows:
\begin{equation}\label{eq:variance.current.exa1}
\begin{aligned}
\mrm{Var}[\jmath]&=\avgl{\int_0^{T}dt\int_0^{T}ds\,\bra{x(t)-\overline{f}}\bra{x(s)-\overline{f}}}\\
&=\int_0^{T}dt\int_0^{T}ds\,\phi(t-s)\\
&=\int_0^{T}dt\int_0^{T}ds\,\bra{A_{+}e^{-c|t-s|}+A_{-}e^{c|t-s|}}\\
&=\frac{2}{c^2}\bras{A_{+}\bra{e^{-cT}+cT-1}+A_{-}\bra{e^{cT}-cT-1}}.
\end{aligned}
\end{equation}
\begin{figure}[t]
	\centering
	\includegraphics[width=8.5cm]{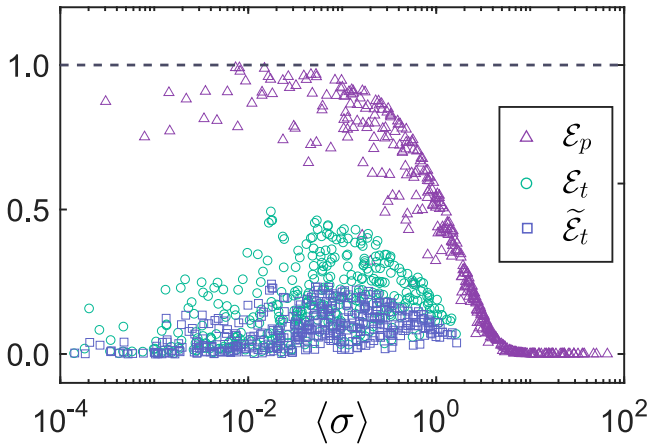}
	\protect\caption{Numerical verification of the TUR in one- and two-dimensional systems. The dashed line represents the saturated TUR. In the one-dimensional system, $\mca{E}_{p}$ is plotted as a function of $\avg{\sigma_{p}}$ with triangular points. The parameter ranges are $a,~f,~D,~\tau,~T\in[0.01,2]$, and $b\in(0,a)$. In the two-dimensional system, $\mca{E}_{t}$ and $\widetilde{\mca{E}}_{t}$ are plotted as functions of $\avg{\sigma_{t}}$ with circular and square points, respectively. The parameter ranges are the same as in the one-dimensional system, except $T\in[0.01,\tau]$. $\mca{E}_{p}\le1,~\mca{E}_{t}\le1$, and $\widetilde{\mca{E}}_{t}\le1$ imply that the derived TUR is satisfied.}\label{fig:examples}
\end{figure}
We define
\begin{equation}
\mca{E}_{p}\equiv\frac{2\avg{\jmath}^2}{\mrm{Var}[\jmath]\bra{e^{\avg{\sigma_{p}}}-1}},
\end{equation}
this should satisfy $\mca{E}_{p}\le 1$.
Using Eq.~\eqref{eq:avg.pos.sigma.smallT} and Eq.~\eqref{eq:variance.current.exa1}, one can numerically evaluate $\mca{E}_{p}$ and verify the TUR for $T\le\tau$.
For the $T>\tau$ case, one can calculate $\avg{\sigma_{p}}$ via Eq.~\eqref{eq:avg.sigma} and obtain
\begin{equation}
\avg{\sigma_{p}}=\frac{Tf^2}{D}+\bra{c\tau+2\frac{A_+e^{-c\tau}-A_{-}}{A_+e^{-c\tau}+A_{-}}}\frac{c\overline{f}^2}{D}.\label{eq:avg.pos.sigma.largeT}
\end{equation}
From Eq.~\eqref{eq:avg.pos.sigma.largeT}, it can be concluded that decreasing the force $f$ or increasing the noise intensity $D$ both result in higher current fluctuation, which is consistent with our intuition.
In the long-time limit $T\to\infty$, we have $\lim_{T\to\infty}T^{-1}\mrm{Var}[\jmath]=\chi_{\jmath}''(0)$, where $\chi_{\jmath}(k)$ is the scaled cumulant generating function defined by $\chi_{\jmath}(k)=\lim_{T\to\infty}T^{-1}\ln\avg{e^{k\jmath}}$.
Using discrete Fourier series, one can obtain $\chi_{\jmath}(k)=k\overline{f}+Dk^2/(a+b)^2$ (see Appendix \ref{app:scgf}).
Therefore, the derived bound can be confirmed for $T\to\infty$ as
\begin{equation}
\frac{\mrm{Var}[\jmath]}{\avg{\jmath}^2}=\frac{2D}{Tf^2}\ge\frac{2}{\avg{\sigma_{p}}}\ge\frac{2}{e^{\avg{\sigma_{p}}}-1}.
\end{equation}

Finally, we run numerical simulations to calculate $\mrm{Var}[\jmath]$ (for $T>\tau$) and verify the bound.
We randomly select parameters $(a,~b,~f,~D,~\tau,~T)$ and repeat the simulations $2\times10^6$ times for each selected parameter setting using time step $\Delta t=10^{-4}$.
We plot $\mca{E}_{p}$ as a function of $\avg{\sigma_{p}}$ as the triangular points in Fig.~\ref{fig:examples}.
The ranges of the parameters are given in the corresponding caption.
As seen, all triangular points are located below the dashed line, which corresponds to the saturated case of the bound; thus, the derived TUR is empirically validated in this system.

Due to the presence of external force $f$, the position symmetry with respect to $0$ is broken in the system.
The degree of broken symmetry is reflected via the quantity $\avg{\sigma_p}$, which is always positive and is a monotonically increasing function of $f$.
Therefore, the derived bound implies that increasing $f$ results in a lower fluctuation.
From a different point of view, since $\jmath=T\overline{f}+\int_0^Tdt\,z$, increasing $f$ enlarges the mean $\avg{\jmath}$ but keeps the variance $\mrm{Var}[\jmath]$ unchanged.
Consequently, the fluctuation of the observable decreases when $f\to\infty$, which is consistent with the conclusion obtained from the TUR.

\subsection{Two-dimensional system}
Here, we consider a simple two-dimensional system with drift force
\begin{equation}
\bm{F}(\bm{x},\bm{x}_\tau)=\begin{bmatrix}
-ax_1+bx_{2,\tau}\\
-ax_2-bx_{1,\tau}
\end{bmatrix}
,
\end{equation}
where $a>b>0$ are the given constants and $x_{i,\tau}\equiv x_i(t-\tau)$.
The noise intensities are set to $D_1=D_2=D$.
This system is manipulated under a parabolic potential with linear delay feedback.
The steady-state distribution $P^{\rm ss}(\bm{x})$ of the system is Gaussian, i.e., $P^{\rm ss}(\bm{x})\propto\exp\bra{-1/2\bm{x}^\top\bm{\Phi}^{-1}\bm{x}}$, because the force is linear.
Here, $\bm{\Phi}$ is the covariance matrix with elements $\Phi_{ij}=\phi_{ij}(0)$, and $\phi_{ij}(z)=\avg{x_i(t)x_j(t+z)}$ denotes the time-correlation function.
The analytical form of this function can be obtained for $|z|\le\tau$ (see Appendix \ref{app:time.corr}).
When $T\le\tau$, $\avg{\sigma_{t}}$ can be calculated using a path integral (see Appendix \ref{app:pat.int})
\begin{equation}
\avg{\sigma_{t}}=\frac{4A_{12}^2\bra{1-e^{-2cT}}}{\bras{(A_{11}^{+})^2+A_{12}^2}e^{-2cT}-\bras{(A_{11}^{-})^2+A_{12}^2}},
\end{equation}
where $c=\sqrt{a^2-b^2}$ and
\begin{equation}
\begin{aligned}
A_{11}^{\pm}&=\frac{D}{2c}\times\frac{(c\pm a)e^{\pm c\tau}}{a\cosh(c\tau)+c\sinh(c\tau)},\\
A_{12}&=\frac{D}{2c}\times\frac{b}{a\cosh(c\tau)+c\sinh(c\tau)}.
\end{aligned}
\end{equation}
As seen, due to the time delay, $\avg{\sigma_t}$ is positive; this implies that the time-reversal symmetry in the system is broken.

Now, we validate the TUR for the following current-type observable
\begin{equation}
\jmath(\bX)=\int_{0}^{T}dt\,\bras{(-ax_1+bx_2)\circ\dot{x}_1+(-ax_2-bx_1)\circ\dot{x}_2}.
\end{equation}
We consider only the $T\le\tau$ case, where $\avg{\sigma_{t}}$ can be analytically obtained.
The effective forces are also linear and can be calculated explicitly (see Appendix \ref{app:eff.force})
\begin{equation}
\overline{F}_1(\bm{x})=-\overline{a}x_1+\overline{b}x_2,~\overline{F}_2(\bm{x})=-\overline{a}x_2-\overline{b}x_1,
\end{equation}
where
\begin{equation}
\begin{aligned}
\overline{a}&=\frac{c\bra{a\cosh(c\tau)+c\sinh(c\tau)}}{a\sinh(c\tau)+c\cosh(c\tau)},\\
\overline{b}&=\frac{bc}{a\sinh(c\tau)+c\cosh(c\tau)}.
\end{aligned}
\end{equation}
The average of the observable is then obtained as
\begin{equation}\label{eq:avg.current.exa2}
\begin{aligned}
\avg{\jmath}&=T\int d\bm{x}\bras{(-ax_1+bx_2)J_1^{\rm ss}(\bm{x})+(-ax_2-bx_1)J_2^{\rm ss}(\bm{x})}\\
&=\frac{2DTb^2}{a\cosh(c\tau)+c\sinh(c\tau)},
\end{aligned}
\end{equation}
which is always positive for an arbitrary delay time.
Equation~\eqref{eq:avg.current.exa2} reveals that increasing $b,~D$, or $T$ leads to a higher average current.
We also consider a non-current observable $\widetilde{\jmath}(\bX)=\mrm{sign}[\jmath(\bX)]$, which represents the sign of the observable $\jmath$; this observable is obviously antisymmetric under time reversal.
We define $\mca{E}_{t}\equiv 2\avg{\jmath}^2/\bras{\mrm{Var}[\jmath]\bra{e^{\avg{\sigma_{t}}}-1}}$ and $\widetilde{\mca{E}}_{t}\equiv 2\avg{\widetilde{\jmath}}^2/\bras{\mrm{Var}[\widetilde{\jmath}]\bra{e^{\avg{\sigma_{t}}}-1}}$, which should satisfy $\mca{E}_{t}\le 1$ and $\widetilde{\mca{E}}_{t}\le 1$.
We run numerical simulations with the same settings as in the one-dimensional system, and plot $\mca{E}_{t}$ and $\widetilde{\mca{E}}_{t}$ as functions of $\avg{\sigma_{t}}$ with circular and square points, respectively, in Fig.~\ref{fig:examples}.
As seen, all circular and square points lie below the dashed line, thus empirically verifying the derived bound.
During the simulation, we have not seen any violation of the inequality $\mrm{Var}[\jmath]/\avg{j}^2\ge 2/\avg{\sigma_t}$.
We conjecture that for continuous-time systems, the fluctuation of arbitrary currents is lower bounded by $2/\avg{\sigma_t}$.
\begin{figure}[t]
	\centering
	\includegraphics[width=8.5cm]{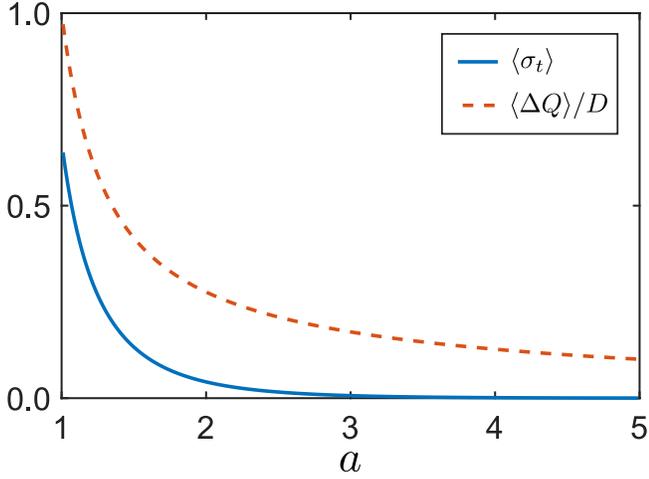}
	\protect\caption{The quantity $\avg{\sigma_t}$ and the average dissipated heat $\avg{\Delta Q}$ in the two-dimensional system. Parameter $a$ is varied from $1$ to $5$, while other parameters are fixed as $b=1$, $D=1$, $T=0.5$, and $\tau=1$.}\label{fig:exa.twodimen}
\end{figure}

Now, we examine the relationship between the term $\avg{\sigma_t}$ and the heat dissipated from the system to the environment.
The heat can be identified as the work done by the system on the environment \cite{Sekimoto.2010,Rosinberg.2015.PRE} and quantified as
\begin{equation}
\Delta Q=\int_0^Tdt\bras{F_1(\bm{x},\bm{x}_\tau)\circ\dot{x}_1+F_2(\bm{x},\bm{x}_\tau)\circ\dot{x}_2}.
\end{equation}
Its average can be calculated analytically as
\begin{equation}\label{eq:heat.ex2}
\begin{aligned}
\avg{\Delta Q}&=T\avg{(-ax_1+bx_{2,\tau})^2+(ax_2+bx_{1,\tau})^2-2aD}\\
&=T\bras{2(a^2+b^2)\phi_{11}(0)+4ab\phi_{12}(\tau)-2aD}\\
&=2DTb^2\times\frac{\cosh(c\tau)}{a\cosh(c\tau)+c\sinh(c\tau)}.
\end{aligned}
\end{equation}
Equation~\eqref{eq:heat.ex2} shows that the average dissipated heat is always nonnegative, i.e., $\avg{\Delta Q}\ge 0$.
We plot $\avg{\sigma_t}$ and $\avg{\Delta Q}/D$ in Fig.~\ref{fig:exa.twodimen} to illustrate how these quantities are related.
We vary the value of $a$, while keeping other parameters unchanged.
As seen, $\avg{\sigma_t}$ and $\avg{\Delta Q}$ show a strong correlation.
When $a$ is increased, both $\avg{\sigma_t}$ and $\avg{\Delta Q}$ decrease.
In particular, $\avg{\Delta Q}$ decreases with order $O(a^{-1})$, while $\avg{\sigma_t}$ declines exponentially.
Indeed, we can prove that $\avg{\sigma_t}\le\avg{\Delta Q}/D$ (see Appendix~\ref{app:ine.proof}).
Consequently, it can be concluded that
\begin{equation}
\frac{\mrm{Var}[\jmath]}{\avg{\jmath}^2}\ge\frac{2}{e^{\avg{\Delta Q}/D}-1},
\end{equation}
which is a direct consequence of the derived bound.
In the region $a\ge 3$, $\avg{\sigma_t}$ is almost zero; this indicates that the system is near equilibrium.
Nonetheless, $\avg{\Delta Q}$ slowly converges to zero due to the time delay.
Therefore, the term $\avg{\sigma_t}$ characterizes the irreversibility in the system better than $\avg{\Delta Q}$ does.

\subsection{Dragged particle in a non-Markovian heat reservoir}

We study a harmonic oscillator of a unit-mass colloidal particle immersed in a heat reservoir at inverse temperature $\beta$ with memory effects \cite{Zon.2003.PRE,Mai.2007.PRE,Paredes.2016.PRE,Ghosh.2017.PA}.
The center of the harmonic potential $U(x,\lambda(t))=k/2(x-\lambda(t))^2$ is dragged by an external protocol $\lambda(t)$.
The dynamics of the system are governed by the following generalized Langevin equation: 
\begin{equation}
\ddot{x}(t)=-\int_0^tds\,\gamma(t-s)\dot{x}(s)-\pp_xU(x,\lambda(t))+\eta(t),\label{eq:harmonic.trap}
\end{equation}
where $\gamma(t)=(\gamma_0/\tau_c)e^{-|t|/\tau_c}$ is the friction memory kernel and $\eta(t)$ is the zero-mean Gaussian colored noise with variance $\avg{\eta(t)\eta(t')}=\beta^{-1}\gamma(t-t')$.
Here, $\tau_c$ denotes the memory time of the heat reservoir and $\gamma_0$ is a positive constant.
It is obvious that the system has distributed time delays.

Hereafter, we consider a time-symmetric protocol given by
\begin{equation}
\lambda(t)=\begin{cases}
	\alpha t, & \text{if}~0\le t< T/2,\\
	\alpha(T-t), & \text{if}~T/2\le t\le T,
\end{cases}
\end{equation}
where $\alpha>0$ is a constant.
This protocol satisfies the condition $\lambda(t)=\lambda(T-t)$.
Suppose that the system is initially in equilibrium, i.e., the initial distribution is of a Maxwell--Boltzmann
type, $P(x,v,0)=\mca{C}\exp\bra{-\beta\bras{v^2/2+U(x,\lambda(0)}}$.
Here, $v\equiv\dot{x}$ is the velocity and $\mca{C}$ is the normalization constant.
Subsequently, the system is coupled with a non-Markovian heat reservoir and driven out of equilibrium by the protocol $\lambda(t)$ during the time interval $[0,T]$.
The heat exchanged between the system and the heat reservoir is defined as
\begin{equation}\label{eq:heat.ex3}
\begin{aligned}
\Delta Q&=\int_0^Tdt\,\bras{\int_0^tds\,\gamma(t-s)\dot{x}(s)-\eta(t)}\circ\dot{x}(t)\\
&=-\int_0^Tdt\,\bras{\ddot{x}(t)+k(x(t)-\lambda(t))}\circ\dot{x}(t).
\end{aligned}
\end{equation}
Because the trajectory $\bXf{0,T}$ is uniquely specified if the noise trajectory $\bm{\eta}\equiv\brab{\eta(t)}_{t=0}^{t=T}$ and the initial condition $\psi(0)\equiv[x(0),v(0)]$ are given, the path probability $\mca{P}(\bXf{0,T}|\psi(0))$ can be expressed by the occurrence probability of the noise trajectory $\bm{\eta}$ as follows:
\begin{equation}
\mca{P}(\bX|\psi(0))\mca{D}\bX=\mca{P}(\bm{\eta})\mca{D}\bm{\eta}.\label{eq:paths.rel}
\end{equation}
Since the noise is Gaussian, the probability of observing trajectory $\bm{\eta}$ is calculated as
\begin{equation}
\mca{P}(\bm{\eta})\propto\exp\bra{-\frac{1}{2}\int_0^Tdt\int_0^Tdt'\,\eta(t)G(t,t')\eta(t')},\label{eq:noise.path.prob}
\end{equation}
where $G(t,t')$ is the inverse of the time-correlation function of the noise and defined as follows:
\begin{equation}
\int_0^Tdt'\,G(t,t')\beta^{-1}\gamma(t'-t'')=\delta(t-t'').
\end{equation}
Plugging Eq.~\eqref{eq:noise.path.prob} into Eq.~\eqref{eq:paths.rel}, the path probability can be readily obtained as
\begin{equation}\label{eq:pat.int.ex3}
\begin{aligned}
&\mca{P}(\bXf{0,T}|\psi(0))=\mca{N}\exp\bigg[-\frac{1}{2}\int_0^Tdt\int_0^Tdt'\,G(t,t')\\
&\times\Big\{\ddot{x}(t)+\int_0^tds\,\gamma(t-s)\dot{x}(s)+k(x(t)-\lambda(t))\Big\}\\
&\times\Big\{\ddot{x}(t')+\int_0^{t'}ds'\,\gamma(t'-s')\dot{x}(s')+k(x(t')-\lambda(t'))\Big\}\bigg],
\end{aligned}
\end{equation}
where $\mca{N}$ is a Jacobian term that is independent of the trajectories.
The quantity $\sigma_t$ can be expressed as
\begin{equation}
\sigma_t=\ln\frac{P(x(0),v(0),0)}{P(x(T),-v(T),0)}+\ln\frac{\mca{P}(\bXf{0,T}|\psi(0))}{\mca{P}(\bXd{0,T}|\psi^\dagger(0))}.\label{eq:sigma.decomp.ex3}
\end{equation}
Here, $\psi^\dagger(0)\equiv[x(T),-v(T)]$.
Using the formula of the path probability in Eq.~\eqref{eq:pat.int.ex3}, we can prove that the second term in the right-hand side of Eq.~\eqref{eq:sigma.decomp.ex3} is equal to the dissipated heat \cite{Ohkuma.2007.JSM}:
\begin{equation}\label{eq:heat.path.ex3}
\ln\frac{\mca{P}(\bXf{0,T}|\psi(0))}{\mca{P}(\bXd{0,T}|\psi^\dagger(0))}=\beta\Delta Q.
\end{equation}
\begin{figure}[t]
	\centering
	\includegraphics[width=8.5cm]{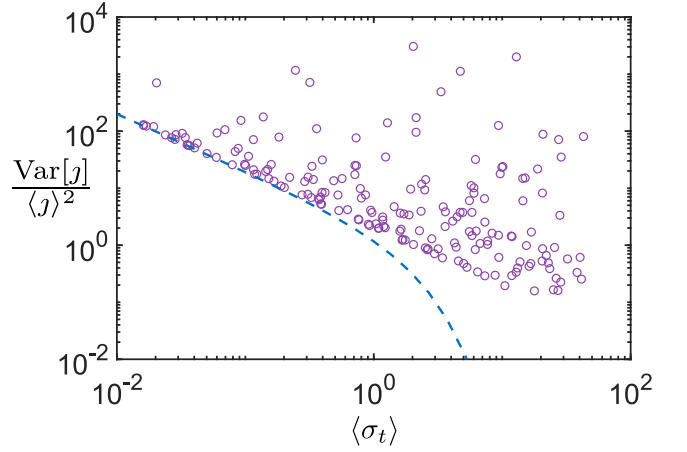}
	\protect\caption{Numerical verification of the TUR in the system of a dragged colloidal particle. The parameter ranges are $\alpha,\beta,\gamma_0,\tau_c,k\in[0.1,2]$, and $T\in[1,10]$. $\mrm{Var}[\jmath]/\avg{\jmath}^2$ is plotted as a function of $\avg{\sigma_t}$ with violet circles. The dashed line represents the derived bound $2/(e^{\avg{\sigma_t}}-1)$. All circular points lie above the line; thus, the derived bound is empirically verified.}\label{fig:exa.nonMarkov}
\end{figure}

We now verify the derived TUR with the current $\jmath(\bX)=\int_0^Tdt\,\dot{x}(t)=x(T)-x(0)$, which expresses the displacement of the particle.
Since this current is odd under time reversal, the inequality $\mrm{Var}[\jmath]/\avg{\jmath}^2\ge2/(e^{\avg{\sigma_t}}-1)$ should be satisfied.
The fluctuation of this current and the derived bound can be calculated analytically.
First, we have that $\avg{x(0)}=\avg{v(0)}=0,~\avg{x(0)^2}=(k\beta)^{-1}$, and $\avg{v(0)^2}=\beta^{-1}$.
The average current is $\avg{\jmath}=\avg{x(T)}-\avg{x(0)}=\avg{x(T)}$.
The variance of the current becomes
\begin{equation}
\mrm{Var}[\jmath]=\avg{x(T)^2}-\avg{x(T)}^2+\avg{x(0)^2}-2\avg{x(0)x(T)}.
\end{equation}
From Eq.~\eqref{eq:heat.ex3}, the average dissipated heat can be calculated as
\begin{equation}\label{eq:avg.heat.ex3}
\begin{aligned}
\avg{\Delta Q}&=\frac{1}{2}\avg{v(0)^2-v(T)^2}+\frac{k}{2}\avg{x(0)^2-x(T)^2}\\
&+k\alpha\avgl{\int_{T/2}^{T}dt\,x(t)-\int_0^{T/2}dt\,x(t)}.
\end{aligned}
\end{equation}
The average of the boundary term in Eq.~\eqref{eq:sigma.decomp.ex3} is
\begin{equation}\label{eq:boundary.term.ex3}
\begin{aligned}
&\avgl{\ln\frac{P(x(0),v(0),0)}{P(x(T),-v(T),0)}}\\
&=\frac{\beta}{2}\avgl{v(T)^2-v(0)^2+k\bra{x(T)^2-x(0)^2}}.
\end{aligned}
\end{equation}
Combining Eqs.~\eqref{eq:avg.heat.ex3} and \eqref{eq:boundary.term.ex3}, we readily obtain
\begin{equation}
\avg{\sigma_t}=k\alpha\beta\avgl{\int_{T/2}^{T}dt\,x(t)-\int_0^{T/2}dt\,x(t)}.
\end{equation}
Using the Laplace transform, analytical forms of $\avg{\jmath}$, $\mrm{Var}[\jmath]$, and $\avg{\sigma_t}$ can be obtained (see Appendix~\ref{app:harmonic.trap} for detailed calculations).
We randomly sample parameters $(\alpha,\beta,\gamma_0,\tau_c,k,T)$ and evaluate $\avg{\jmath}$, $\mrm{Var}[\jmath]$, and $\avg{\sigma_t}$ using Eq.~\eqref{eq:analytical.forms.ex3}.
The parameter ranges are given in the caption of Fig.~\ref{fig:exa.nonMarkov}.
As seen in this figure, the derived bound is satisfied for all parameter settings.
In the region $\avg{\sigma_t}<1$, some circular points touch the line, which implies that the derived bound is attainable when the system is near equilibrium.
As in the example of the two-dimensional system, we find that $\mrm{Var}[\jmath]/\avg{\jmath}^2\ge2/\avg{\sigma_t}$ is satisfied for all selected parameters.
This evidence strengthens the conjecture made in the preceding example.

We next consider a physical interpretation of the term $\avg{\sigma_t}$ in this system.
From Eqs.~\eqref{eq:sigma.decomp.ex3} and \eqref{eq:heat.path.ex3}, we have
\begin{equation}
\avg{\sigma_t}=\avgl{\ln\frac{P(\psi(0),0)}{P(\psi^\dagger(0),0)}}+\beta\avg{\Delta Q}.
\end{equation}
As seen, there are two contributions in $\avg{\sigma_t}$, the boundary term $\avgl{\ln{P(\psi(0),0)}/{P(\psi^\dagger(0),0)}}$ and the dissipated heat $\beta\avg{\Delta Q}$.
Neglecting this boundary value, one can approximate $\avg{\sigma_t}\approx\beta\avg{\Delta Q}$.
This implies that $\avg{\sigma_t}$ can be interpreted as the average dissipated heat in the system.
We note that for general cases, i.e., the protocol is time asymmetric, this is not the case.

\section{Conclusion}
In summary, we derived the TUR for the time-delayed systems.
We provided two bounds on the relative fluctuations of general dynamical observables that are antisymmetric under conjugate operations.
For observables that are antisymmetric under time reversal, the fluctuation is lower bounded by $2/(e^{\avg{\sigma_{t}}}-1)$, where $\avg{\sigma_{t}}$ can be considered a generalization of the total entropy production.
On the other hand, the fluctuation of observables that are odd under position reversal is constrained by $\avg{\sigma_{p}}$, which reflects the degree of position-symmetry breaking in the system.
These results hold for an arbitrary observation time.
Because it is not necessary to know the underlying dynamics of the systems, the derived TUR holds for a large class of continuous- and discrete-time systems.
The bound can be used as a tool to estimate a hidden thermodynamic quantity in real-world systems that involve time delays from finite-time experimental data.

From the results in the numerical experiment, we conjectured that the fluctuation of arbitrary time-integrated currents in continuous-time systems is bounded from below by the reciprocal of $\avg{\sigma_t}$.
Proving this inequality would substantially improve the bound and requires further investigation.

\section*{Acknowledgment}
This work was supported by MEXT KAKENHI Grant No.~JP16K00325, JP19K12153.

\appendix
\section{Scaled cumulant generating function of observables}\label{app:scgf}
Here we calculate the scaled cumulant generating function (SCGF) of the observable $\jmath(\bX)=\int_0^Tdt\,x$ in the long-time limit $T\to\infty$.
Note that $\jmath=T\overline{f}+\int_0^Tdt\,z$.
By imposing periodic boundary conditions on the trajectories, $z(t)$ can be expanded in a discrete Fourier series \cite{Rosinberg.2016.EPL} as 
\begin{equation}
z(t)=\sum_{n=-\infty}^{\infty}z_ne^{-i\omega_nt},\label{eq:discrete.Fourier}
\end{equation}
where the coefficient $z_n$ can be calculated via inverse transforms
\begin{equation}
z_n=\frac{1}{T}\int_0^Tdt\,z(t)e^{i\omega_nt},
\end{equation}
where $\omega_n=2\pi n/T$.
By substituting Eq.~\eqref{eq:discrete.Fourier} into the Langevin equation, we obtain
\begin{equation}
(a+be^{i\omega_n\tau}-i\omega_n)z_n=\xi_n,
\end{equation}
with $\avg{\xi_n\xi_m}=2D/T\delta_{n,-m}$.
The current $\jmath$ can then be expressed as $\jmath=T\overline{f}+Tz_0=T\overline{f}+T\xi_0/(a+b)$.
Substituting $\jmath$ into the definition of the SCGF, we obtain
\begin{widetext}
\begin{align}
\chi_{\jmath}(k)&=\lim_{T\to\infty}T^{-1}\ln\avgl{\exp\bra{kT(\overline{f}+\xi_0/(a+b))}}\nonumber\\
&=k\overline{f}+\lim_{T\to\infty}T^{-1}\ln\bra{\int_{-\infty}^{\infty} d\xi_0\,P(\xi_0)\exp\bras{kT\xi_0/(a+b)}},\label{eq:discrete.SCGF}
\end{align}
\end{widetext}
where $P(\xi_0)=\sqrt{T/(4\pi D)}\exp\bras{-T\xi_0^2/(4D)}$.
Taking the integration in Eq.~\eqref{eq:discrete.SCGF}, we get $\chi_{\jmath}(k)=k\overline{f}+Dk^2/(a+b)^2$.

\section{Detailed derivations in the two-dimensional system}
\subsection{Time-correlation function}\label{app:time.corr}
Here, we calculate the stationary time-correlation function $\phi_{ij}(z)=\avg{x_i(t)x_j(t+z)}$.
Using the same method as in Ref.~\cite{Frank.2003.PRE} for arbitrary $z\ge 0$, we have
\begin{equation}
\begin{aligned}
\dz\phi_{11}(z)=-a\phi_{11}(z)+b\phi_{21}(\tau-z)+\avg{x_1(t)\xi_1(t+z)},\\
\dz\phi_{12}(z)=-a\phi_{12}(z)-b\phi_{11}(\tau-z)+\avg{x_1(t)\xi_2(t+z)},\\
\dz\phi_{21}(z)=-a\phi_{21}(z)+b\phi_{22}(\tau-z)+\avg{x_2(t)\xi_1(t+z)},\\
\dz\phi_{22}(z)=-a\phi_{22}(z)-b\phi_{12}(\tau-z)+\avg{x_2(t)\xi_2(t+z)}.
\end{aligned}
\end{equation}
From the Fokker-Planck equation, we have
\begin{equation}
0=\frac{d}{dt}\avg{x_1^2}=-2a\phi_{11}(0)+2b\phi_{21}(\tau)+2D.\label{eq:correlation.eq1}
\end{equation}
On the other hand, from Langevin equation, we also obtain
\begin{equation}
0=\frac{d}{dt}\avg{x_1^2}=-2a\phi_{11}(0)+2b\phi_{21}(\tau)+2\avg{x_1(t)\xi_1(t)}.\label{eq:correlation.eq2}
\end{equation}
Comparing Eq.~\eqref{eq:correlation.eq1} and Eq.~\eqref{eq:correlation.eq2}, we obtain the relation $\avg{x_1(t)\xi_1(t)}=D$.
Similarly, we also get $\avg{x_2(t)\xi_2(t)}=D,~\avg{x_1(t)\xi_2(t)}+\avg{x_2(t)\xi_1(t)}=0$.
Because the noise is irrelevant to the past states of the system, we have $\avg{x_i(t)\xi_j(t+z)}=0,~\forall z>0$.
Using the Fourier transform $\bm{g}(\omega)=\int_{-\infty}^{\infty}dt\,e^{i\omega t}\bm{g}(t)$ for an arbitrary function $\bm{g}(t)$, we obtain the relation that $\bm{x}(\omega)=\bm{H}(\omega)\bm{\xi}(\omega)$. Here, $\bm{H}(\omega)$ is a response function matrix in the frequency domain, given by
\begin{equation}
\bm{H}(\omega)=\frac{1}{(a-i\omega)^2+b^2e^{i2\omega\tau}}\times\begin{pmatrix}
a-i\omega & be^{i\omega\tau}\\
-be^{i\omega\tau} & a-i\omega
\end{pmatrix}.
\end{equation}
The time-correlation function can be calculated via an inverse Fourier transform of the spectral density $\bm{S}(\omega)$ given by
\begin{equation}
\bm{S}(\omega)=2\bm{H}(\omega)\bm{D}\bm{H}^*(\omega),
\end{equation}
where $\bm{D}={\rm diag}(D,D)\in\mbb{R}^{2\times2}$ and $\bm{H}^*$ is the complex conjugate transpose of $\bm{H}$.
Since $S_{11}(\omega)=S_{22}(\omega),~S_{12}(\omega)+S_{21}(\omega)=0$, we readily obtain
\begin{equation}
\phi_{11}(z)=\phi_{22}(z),~\phi_{12}(z)+\phi_{21}(z)=0.\label{eq:correlation.relations}
\end{equation}
Using the relations in Eq.~\eqref{eq:correlation.relations}, we obtain that for $0\le z\le\tau$
\begin{equation}\label{eq:phi11.solution}
\dzt\phi_{11}(z)=(a^2-b^2)\phi_{11}(z).
\end{equation}
The solution of time-correlation function $\phi_{11}(z)$ in Eq.~\eqref{eq:phi11.solution} has the following form:
\begin{equation}
\phi_{11}(z)=\alpha\cosh(cz)+\beta\sinh(cz),
\end{equation}
where $c=\sqrt{a^2-b^2}$ and $\alpha,~\beta$ are constants determined via the conditions:
\begin{equation}
\left.\dz\phi_{11}(z)\right|_{z\to 0}=-D,\quad \left.\phi_{12}(z)\right|_{z\to 0}=0.
\end{equation}
Finally, we obtain that for $0\le z\le\tau$
\begin{align}
\phi_{11}(z)&=\phi_{22}(z)=A_{11}^{+}e^{-cz}+A_{11}^{-}e^{cz},\\
\phi_{12}(z)&=-\phi_{21}(z)=A_{12}\bra{e^{-cz}-e^{cz}},
\end{align}
where
\begin{align}
A_{11}^{\pm}&=\frac{D}{2c}\times\frac{(c\pm a)e^{\pm c\tau}}{a\cosh(c\tau)+c\sinh(c\tau)},\\
A_{12}&=\frac{D}{2c}\times\frac{b}{a\cosh(c\tau)+c\sinh(c\tau)}.
\end{align}
Because $\phi_{11}(z)$ is an even function and $\phi_{12}(z)$ is an odd function, we readily obtain that for $|z|\le\tau$,
\begin{equation}
\begin{aligned}
\phi_{11}(z)&=\phi_{22}(z)=A_{11}^{+}e^{-c|z|}+A_{11}^{-}e^{c|z|},\\
\phi_{12}(z)&=-\phi_{21}(z)=A_{12}\bra{e^{-cz}-e^{cz}}.
\end{aligned}
\end{equation}

\begin{widetext}
\subsection{Path integral}\label{app:pat.int}
Because the process is Gaussian, the path probability is given by
\begin{equation}\label{eq:path.integral.Gaussian.process}
\mca{P}(\bX)\propto\exp\bra{-\frac{1}{2}\int_0^Tdt\int_0^Tdt'\,[x_1(t)~x_2(t)]\begin{bmatrix}
	\Gamma_{11}(t,t') & \Gamma_{12}(t,t')\\
	\Gamma_{21}(t,t') & \Gamma_{22}(t,t')\\
	\end{bmatrix}
	\begin{bmatrix}
	x_1(t')\\
	x_2(t')
	\end{bmatrix}},
\end{equation}
where $\Gamma_{ij}(t,t')$ is the inverse of the stationary time-correlation function $\phi_{ij}(z)$ defined via the following relation:
\begin{equation}\label{eq:inverse.matrix.condition}
\int_0^Tds\begin{bmatrix}
\phi_{11}(t-s) & \phi_{12}(t-s)\\
\phi_{21}(t-s) & \phi_{22}(t-s)
\end{bmatrix}
\begin{bmatrix}
\Gamma_{11}(s,t') & \Gamma_{12}(s,t')\\
\Gamma_{21}(s,t') & \Gamma_{22}(s,t')
\end{bmatrix}
=\begin{bmatrix}
\delta(t-t') & 0\\
0 & \delta(t-t')
\end{bmatrix}.
\end{equation}
Now, we discretize the problem and take the continuum limit at the end.
We divide the time interval $[0,T]$ into $N$ equipartitioned intervals with a time step $\epsilon=T/N$, where $t_k=k\epsilon~(k=0,\dots,N)$ and $x_1^k=x_1(t_k),~x_2^k=x_2(t_k)$ (superscripts denote points in a temporal sequence).
Equation~\eqref{eq:path.integral.Gaussian.process} then reads
\begin{equation}
\mca{P}(x_1^0,x_2^0,t_0;\dots;x_1^N,x_2^N,t_N)\propto\exp\bra{-\frac{1}{2}\sum_{i,j}\bras{x_1^i\Gamma_{11}^{ij}x_1^j+x_1^i\Gamma_{12}^{ij}x_2^j+x_2^i\Gamma_{21}^{ij}x_1^j+x_2^i\Gamma_{22}^{ij}x_2^j}},\label{eq:discrete.path.integral}
\end{equation}
and Eq.~\eqref{eq:inverse.matrix.condition} corresponds to the following equation:
\begin{equation}
\sum_{p=1}^{2}\sum_{j=0}^N\phi_{mp}^{ij}\Gamma_{pn}^{jk}=\delta_{mn}\delta_{ik},
\end{equation}
where $\phi_{mp}^{ij}\equiv\phi_{mp}(t_j-t_i)$.
The matrices $\Gamma_{mn}~(1\le m,~n\le 2)$ can be analytically calculated and have the following form:
\begin{equation}\label{eq:inverse.matrix.form}
\begin{aligned}
\Gamma_{11}&=\Gamma_{22},~\Gamma_{12}=-\Gamma_{21},\\
\Gamma_{11}^{0N}&=\Gamma_{11}^{N0}=\frac{e^{-Nc\epsilon}\left(A_{11}^{+}A_{11}^{-}+A_{12}^2\right)}{(A_{11}^{+}-A_{11}^{-})\left((A_{11}^{-})^2+A_{12}^2-\left((A_{11}^{+})^2+A_{12}^2\right)e^{-2Nc\epsilon}\right)},\\
\Gamma_{11}^{ij}&=0,~\forall~1<|i-j|<N,\\
\Gamma_{11}^{ij}&=\frac{-e^{-c\epsilon}}{(A_{11}^{+}-A_{11}^{-})(1-e^{-2c\epsilon})},~\forall~|i-j|=1,\\
\Gamma_{11}^{ii}&=\frac{1+e^{-2c\epsilon}}{(A_{11}^{+}-A_{11}^{-})(1-e^{-2c\epsilon})},~\forall~0<i<N,\\
\Gamma_{11}^{00}&=\Gamma_{11}^{NN}=\frac{e^{-2c\epsilon}\bras{(A_{11}^{-})^2+A_{12}^2-\bra{(A_{11}^{+})^2+A_{12}^2}e^{-2(N-1)c\epsilon}}}{(A_{11}^{+}-A_{11}^{-})\bra{1-e^{-2c\epsilon}}\bras{(A_{11}^{-})^2+A_{12}^2-((A_{11}^{+})^2+A_{12}^2)e^{-2Nc\epsilon}}},\\
\Gamma_{12}^{0N}&=-\Gamma_{12}^{N0}=\frac{-A_{12}e^{-Nc\epsilon}}{(A_{11}^{-})^2+A_{12}^2-((A_{11}^{+})^2+A_{12}^2)e^{-2Nc\epsilon}},\\
\Gamma_{12}^{ij}&=0,~\forall~|i-j|\neq N.
\end{aligned}
\end{equation}
Using the result in Eq.~\eqref{eq:inverse.matrix.form}, the quadratic form in Eq.~\eqref{eq:discrete.path.integral} can be obtained explicitly as
\begin{equation}
\begin{aligned}
&\sum_{i,j}\bras{x_1^i\Gamma_{11}^{ij}x_1^j+x_1^i\Gamma_{12}^{ij}x_2^j+x_2^i\Gamma_{21}^{ij}x_1^j+x_2^i\Gamma_{22}^{ij}x_2^j}\\
&=\frac{1}{A_{11}^{+}-A_{11}^{-}}\bra{\sum_{i=1}^{2}\sum_{k=1}^N\frac{(x_i^k-e^{-c\epsilon}x_i^{k-1})^2}{1-e^{-2c\epsilon}}-\frac{1}{\Omega_T}\sum_{i=1}^{2}\bras{A_{12}^2\bra{e^{-Nc\epsilon}x_i^0-x_i^N}^2+\bra{A_{11}^{+}e^{-Nc\epsilon}x_i^0-A_{11}^{-}x_i^N}^2}}\\
&-\frac{2A_{12}e^{-Nc\epsilon}}{\Omega_T}\bra{x_1^0x_2^N-x_1^Nx_2^0},
\end{aligned}
\end{equation}
where $\Omega_T=(A_{11}^{-})^2+A_{12}^2-((A_{11}^{+})^2+A_{12}^2)e^{-2cT}$.
Taking the continuum limit $\epsilon\to 0,~N\to\infty$, with $N\epsilon=T$ gives \cite{Rosinberg.2015.PRE}
\begin{equation}
\lim_{\epsilon\to 0}\sum_{k=1}^N\frac{(x_i^k-e^{-c\epsilon}x_i^{k-1})^2}{1-e^{-2c\epsilon}}=\frac{1}{2c}\int_0^Tdt\,\bra{\dot{x}_i(t)+cx_i(t)}^2.
\end{equation}
Finally, we obtain the expression of the path probability for $T\le\tau$:
\begin{equation}
\begin{aligned}
\mca{P}(\bX)&\propto\exp\bra{-\sum_{i=1}^2\int_0^Tdt\,\frac{\bras{\dot{x}_i(t)+cx_i(t)}^2}{4D}}\\
&\times\exp\bra{\frac{c}{2D\Omega_T}\sum_{i=1}^2\brab{A_{12}^2\bras{e^{-cT}x_i(0)-x_i(T)}^2+\bras{A_{11}^{+}e^{-cT}x_i(0)-A_{11}^{-}x_i(T)}^2}}\\
&\times\exp\bra{\frac{A_{12}e^{-cT}}{\Omega_T}\bras{x_1(0)x_2(T)-x_1(T)x_2(0)}}.
\end{aligned}
\end{equation}

\subsection{Analytical form of the effective forces}\label{app:eff.force}
We calculate the analytical form of the effective force $\overline{F}_i(\bm{x})$ from its definition.
We note that $\overline{F}_i(\bm{x})$ cannot be completely determined from the steady-state FPE, i.e., $\sum_{i=1}^2\partial_{x_i}[\overline{F}_i(\bm{x})P(\bm{x},t)-D\partial_{x_i}P(\bm{x},t)]=0$.
Specifically, if the effective force takes the form $\overline{F}_i(\bm{x})=\sum_{j=1}^2\gamma_{ij}x_j$, then one obtains $\gamma_{11}=\gamma_{22}=-D/\phi_{11}(0),~\gamma_{12}+\gamma_{21}=0$.
Here, we use the path integral to calculate $\overline{F}_i(\bm{x})$.
From the definition, we have
\begin{equation}
\begin{aligned}
\overline{F}_i(\bm{v})&=\int d\bm{u}\,F_i(\bm{v},\bm{u})P(\bm{u},t-\tau|\bm{v},t)=\int d\bm{u}\,F_i(\bm{v},\bm{u})P(\bm{v},t;\bm{u},t-\tau)/P(\bm{v},t)\\
&=\int d\bm{u}\,\frac{F_i(\bm{v},\bm{u})}{P(\bm{v},t)}\int_{\bm{u}}^{\bm{v}}\mca{D}\bX\,\mca{P}(\bX),
\end{aligned}
\end{equation}
where the integration is taken over all trajectories $\bX$ that start from $\bm{u}$ at time $t-\tau$ and end at $\bm{v}$ at time $t$.
The first term in the path probability can be simplified further using the well-known expression of the transition probability for Smoluchowski processes \cite{Risken.1989,Rosinberg.2015.PRE}
\begin{equation}
\int_{x(0)}^{x(\tau)}\mca{D}\bX\,\exp\bra{-\int_{0}^{\tau}dt\,\frac{\bras{\dot{x}(t)+cx(t)}^2}{4D}}\propto\exp\bra{-\frac{c}{2D}\frac{\bras{x(\tau)-x(0)e^{-c\tau}}^2}{1-e^{-2c\tau}}}.
\end{equation}
Consequently, we obtain
\begin{equation}
\overline{F}_i(\bm{v})=\int d\bm{u}\frac{F_i(\bm{v},\bm{u})}{P(\bm{v},t)}G(\bm{v},\bm{u}),\label{eq:effective.force.integration}
\end{equation}
where
\begin{equation}
G(\bm{v},\bm{u})\propto\exp\bra{-\frac{c}{2D}\frac{\norm{\bm{v}-\bm{u}e^{-c\tau}}^2}{1-e^{-2c\tau}}+\frac{c}{2D\Omega_\tau}\bra{A_{12}^2\norm{e^{-c\tau}\bm{u}-\bm{v}}^2+\norm{A_{11}^{+}e^{-c\tau}\bm{u}-A_{11}^{-}\bm{v}}^2}+\frac{A_{12}e^{-c\tau}}{\Omega_\tau}\bras{u_1v_2-u_2v_1}}.
\end{equation}
Taking the integration in Eq.~\eqref{eq:effective.force.integration}, we obtain
\begin{equation}
\begin{aligned}
\overline{F}_1(\bm{x})&=-\frac{c\bra{a\cosh(c\tau)+c\sinh(c\tau)}}{a\sinh(c\tau)+c\cosh(c\tau)}x_1+\frac{bc}{a\sinh(c\tau)+c\cosh(c\tau)}x_2,\\
\overline{F}_2(\bm{x})&=-\frac{c\bra{a\cosh(c\tau)+c\sinh(c\tau)}}{a\sinh(c\tau)+c\cosh(c\tau)}x_2-\frac{bc}{a\sinh(c\tau)+c\cosh(c\tau)}x_1.
\end{aligned}
\end{equation}
\end{widetext}

\subsection{Proof of inequality $\avg{\sigma_t}\le\avg{\Delta Q}/D$}\label{app:ine.proof}

Here we provide a proof of $\avg{\sigma_t}\le\avg{\Delta Q}/D$ for $T\le\tau$.
By simple calculations, we can show that
\begin{equation}
\avg{\sigma_t}=\frac{4b^2(1-e^{-2cT})}{\bras{b^2+(c+a)^2e^{2c\tau}}e^{-2cT}-\bras{b^2+(c-a)^2e^{-2c\tau}}}.
\end{equation}
For convenience, we define $U\equiv b^2+(c+a)^2e^{2c\tau}$ and $V\equiv b^2+(c-a)^2e^{-2c\tau}$.
Then, $\avg{\sigma_t}$ can be rewritten as
\begin{equation}
\avg{\sigma_t}=\frac{4b^2(1-e^{-2cT})}{Ue^{-2cT}-V}.
\end{equation}
From Eq.~\eqref{eq:heat.ex2}, we also have
\begin{equation}
\frac{\avg{\Delta Q}}{D}=2Tb^2\times\frac{\cosh(c\tau)}{a\cosh(c\tau)+c\sinh(c\tau)}.
\end{equation}
Therefore, $\avg{\sigma_t}\le\avg{Q}/D$ is equivalent to
\begin{equation}
\frac{2(1-e^{-2cT})}{Ue^{-2cT}-V}\le T\times\frac{\cosh(c\tau)}{a\cosh(c\tau)+c\sinh(c\tau)}.\label{eq:ine.tmp1}
\end{equation}
To prove inequality \eqref{eq:ine.tmp1}, we will show that
\begin{equation}
f(T)\le \frac{\cosh(c\tau)}{a\cosh(c\tau)+c\sinh(c\tau)},\label{eq:ine.tmp2}
\end{equation}
where 
\begin{equation}
f(T)=\frac{2(1-e^{-2cT})}{T(Ue^{-2cT}-V)}.
\end{equation}
First, taking the derivative of $f(T)$, we have
\begin{equation}
\frac{df(T)}{dT}=\frac{e^{2cT}\bra{Ue^{-2cT}+Ve^{2cT}-\bras{U+V+2cT(V-U)}}}{T^2\bra{U-Ve^{2cT}}^2}.
\end{equation}
Since $e^{z}\ge 1+z$, $\forall z\in\mbb{R}$, we have $Ue^{-2cT}+Ve^{2cT}-\bras{U+V+2cT(V-U)}\ge0$; thus, $df(T)/dT\ge 0$.
Consequently, we obtain $f(T)\le f(\tau)$ for all $T\le\tau$.
Therefore, to prove Eq.~\eqref{eq:ine.tmp2}, we need to prove only that
\begin{equation}
f(\tau)\le\frac{\cosh(c\tau)}{a\cosh(c\tau)+c\sinh(c\tau)}.\label{eq:ine.tmp3}
\end{equation}
Inequality \eqref{eq:ine.tmp3} is equivalent to
\begin{equation}
e^{c\tau}-e^{-c\tau}\le{c\tau}\bra{e^{c\tau}+e^{-c\tau}},\label{eq:ine.tmp4}
\end{equation}
which is always satisfied because for all $z\ge 0$,
\begin{equation}
\begin{aligned}
&\frac{d}{dz}\bras{z(e^{z}+e^{-z})-(e^{z}-e^{-z})}=z(e^{z}-e^{-z})\ge 0,\\
&\left.\bras{z(e^{z}+e^{-z})-(e^{z}-e^{-z})}\right|_{z=0}=0.
\end{aligned}
\end{equation}
This implies that $\avg{\sigma_t}\le\avg{\Delta Q}/D$ for $T\le\tau$.

\subsection{Analytical calculations in the dragged colloidal particle model}\label{app:harmonic.trap}

Applying the Laplace transform to Eq.~\eqref{eq:harmonic.trap}, we obtain
\begin{equation}
\begin{aligned}
&s^2\tilde{x}(s)-sx(0)-v(0)+\tilde{\gamma}(s)(s\tilde{x}(s)-x(0))+k\tilde{x}(s)\\
&=k\tilde{\lambda}(s)+\tilde{\eta}(s).
\end{aligned}
\end{equation}
Here, $\tilde{f}(s)=\int_0^{\infty}dt\,f(t)e^{-st}$ is the Laplace transform of an arbitrary function $f(t)$.
The solution to Eq.~\eqref{eq:harmonic.trap} is
\begin{equation}
x(t)=H(t)x(0)+G(t)v(0)+\int_0^tdt'\,G(t-t')\bras{k\lambda(t')+\eta(t')},\label{eq:x.sol}
\end{equation}
where $H(t)$ and $G(t)$ are given by
\begin{align}
H(t)&=\mca{L}^{-1}\brab{\frac{\tilde{\gamma}(s)+s}{s^2+s\tilde{\gamma}(s)+k}},\\
G(t)&=\mca{L}^{-1}\brab{\frac{1}{s^2+s\tilde{\gamma}(s)+k}}.
\end{align}
Here, $\mca{L}^{-1}\set{\cdot}$ denotes the inverse Laplace transform.
We note that $H(t)$ and $G(t)$ satisfy the following differential equations:
\begin{align}
\dot{H}(t)&=-kG(t),\\
\dot{G}(t)&=H(t)-\int_0^tdt'\,\gamma(t-t')G(t'),
\end{align}
with initial conditions $H(0)=\dot{G}(0)=1$ and $G(0)=\dot{H}(0)=0$.
Now, we calculate $G(t)$.
Since $\tilde{\gamma}(s)=\gamma_0/(s\tau_c+1)$, we have
\begin{equation}
G(t)=\mca{L}^{-1}\brab{\frac{s+a}{s^3+as^2+bs+c}},
\end{equation}
where $a=1/\tau_c$, $b=k+\gamma_0/\tau_c$, and $c=k/\tau_c$.
The roots of the polynomial $s^3+as^2+bs+c$ are characterized by the following quantity:
\begin{equation}
Q=-\frac{a^2b^2}{108}+\frac{b^3}{27}+\frac{a^3c}{27}-\frac{abc}{6}+\frac{c^2}{4}.
\end{equation}
In particular, the polynomial has three real roots when $Q<0$, one real root and two complex roots when $Q>0$, and a multiple root when $Q=0$.
Here, we consider only the case $Q>0$ (i.e., the underdamped case).
The denominator can be decomposed as
\begin{equation}
s^3+as^2+bs+c=(s+p)(s+q+i\omega)(s+q-i\omega),
\end{equation}
where
\begin{equation}
\begin{aligned}
p=\frac{a}{3}-A-B,~q=\frac{a}{3}+\frac{A+B}{2},~\omega=\frac{\sqrt{3}}{2}(A-B).
\end{aligned}
\end{equation}
Here, constants $A$ and $B$ are given by
\begin{equation}
\begin{aligned}
A&=\sqrt[3]{-\frac{a^3}{27}+\frac{ab}{6}-\frac{c}{2}+\sqrt{Q}},\\
B&=\sqrt[3]{-\frac{a^3}{27}+\frac{ab}{6}-\frac{c}{2}-\sqrt{Q}}.
\end{aligned}
\end{equation}
Then, $G(t)$ and $H(t)$ can be obtained as
\begin{equation}
\begin{aligned}
G(t)&=c_1e^{-pt}+c_2e^{-qt}\sin(\omega t+\phi),\\
H(t)&=1-k\int_0^tdt'\,G(t'),
\end{aligned}
\end{equation}
where
\begin{align}
c_1&=\frac{a-p}{(p-q)^2+\omega^2},~c_2=\frac{1}{\omega}\sqrt{\frac{(a-q)^2+\omega^2}{(p-q)^2+\omega^2}},\\
\sin\phi&=\frac{\omega(p-a)}{\sqrt{((a-q)^2+\omega^2)((p-q)^2+\omega^2)}},\\
\cos\phi&=\frac{(a-q)(p-q)+\omega^2}{\sqrt{((a-q)^2+\omega^2)((p-q)^2+\omega^2)}}.
\end{align}
Once the functions $G(t)$ and $H(t)$ are obtained, the fluctuation of the current and the derived bound can be calculated immediately.
From Eq.~\eqref{eq:x.sol}, we have
\begin{align}
\avg{x(t)}&=k\int_0^tdt'\,G(t-t')\lambda(t'),\\
\avg{x(0)x(t)}&=H(t)\avg{x(0)^2}.
\end{align}
Consequently, we obtain the following results: 
\begin{widetext}
\begin{equation}\label{eq:analytical.forms.ex3}
\begin{aligned}
H(t)&=1-\frac{kc_1(1-e^{-pt})}{p}-\frac{kc_2\bra{\omega\cos\phi+q\sin\phi-e^{-qt}\bras{\omega\cos(\omega t+\phi)+q\sin(\omega t+\phi)}}}{q^2+\omega^2},\\
\avg{x(T)}&=k\alpha\bras{\int_0^{T/2}dt\,G(T-t)t+\int_{T/2}^Tdt\,G(T-t)(T-t)}\\
&=\frac{k\alpha c_1(e^{-pT/2}-1)^2}{p^2}+\frac{k\alpha c_2}{(q^2+\omega^2)^2}\Big[2q\omega\bra{\cos\phi-2e^{-qT/2}\cos(\omega T/2+\phi)+e^{-qT}\cos(\omega T+\phi)}\\
&+(q^2-\omega^2)\bra{\sin\phi-2e^{-qT/2}\sin(\omega T/2+\phi)+e^{-qT}\sin(\omega T+\phi)}\Big],\\
\avg{x(T)^2}&=(k\beta)^{-1}H(T)^2+\beta^{-1}G(T)^2+k^2\bra{\int_0^Tdt\,G(T-t)\lambda(t)}^2+\beta^{-1}\frac{\gamma_0}{\tau_c}\int_0^Tdt\int_0^Tdt'\,G(T-t)G(T-t')e^{-\frac{|t-t'|}{\tau_c}},\\
\avg{\sigma_t}&=k^2\alpha\beta\bras{\int_{T/2}^Tdt\int_0^tdt'\,G(t-t')\lambda(t')-\int_0^{T/2}dt\int_0^tdt'\,G(t-t')\lambda(t')}.
\end{aligned}
\end{equation}
\end{widetext}


\begin{thebibliography}{67}%
	\makeatletter
	\providecommand \@ifxundefined [1]{%
		\@ifx{#1\undefined}
	}%
	\providecommand \@ifnum [1]{%
		\ifnum #1\expandafter \@firstoftwo
		\else \expandafter \@secondoftwo
		\fi
	}%
	\providecommand \@ifx [1]{%
		\ifx #1\expandafter \@firstoftwo
		\else \expandafter \@secondoftwo
		\fi
	}%
	\providecommand \natexlab [1]{#1}%
	\providecommand \enquote  [1]{``#1''}%
	\providecommand \bibnamefont  [1]{#1}%
	\providecommand \bibfnamefont [1]{#1}%
	\providecommand \citenamefont [1]{#1}%
	\providecommand \href@noop [0]{\@secondoftwo}%
	\providecommand \href [0]{\begingroup \@sanitize@url \@href}%
	\providecommand \@href[1]{\@@startlink{#1}\@@href}%
	\providecommand \@@href[1]{\endgroup#1\@@endlink}%
	\providecommand \@sanitize@url [0]{\catcode `\\12\catcode `\$12\catcode
		`\&12\catcode `\#12\catcode `\^12\catcode `\_12\catcode `\%12\relax}%
	\providecommand \@@startlink[1]{}%
	\providecommand \@@endlink[0]{}%
	\providecommand \url  [0]{\begingroup\@sanitize@url \@url }%
	\providecommand \@url [1]{\endgroup\@href {#1}{\urlprefix }}%
	\providecommand \urlprefix  [0]{URL }%
	\providecommand \Eprint [0]{\href }%
	\providecommand \doibase [0]{http://dx.doi.org/}%
	\providecommand \selectlanguage [0]{\@gobble}%
	\providecommand \bibinfo  [0]{\@secondoftwo}%
	\providecommand \bibfield  [0]{\@secondoftwo}%
	\providecommand \translation [1]{[#1]}%
	\providecommand \BibitemOpen [0]{}%
	\providecommand \bibitemStop [0]{}%
	\providecommand \bibitemNoStop [0]{.\EOS\space}%
	\providecommand \EOS [0]{\spacefactor3000\relax}%
	\providecommand \BibitemShut  [1]{\csname bibitem#1\endcsname}%
	\let\auto@bib@innerbib\@empty
	\bibitem [{\citenamefont {Sekimoto}(1998)}]{Sekimoto.1998.PTPS}%
	\BibitemOpen
	\bibfield  {author} {\bibinfo {author} {\bibfnamefont {K.}~\bibnamefont
			{Sekimoto}},\ }\href {\doibase 10.1143/PTPS.130.17} {\bibfield  {journal}
		{\bibinfo  {journal} {Prog. Theor. Phys. Supp.}\ }\textbf {\bibinfo {volume}
			{130}},\ \bibinfo {pages} {17} (\bibinfo {year} {1998})}\BibitemShut
	{NoStop}%
	\bibitem [{\citenamefont {Evans}\ and\ \citenamefont
		{Searles}(2002)}]{Denis.2002.AP}%
	\BibitemOpen
	\bibfield  {author} {\bibinfo {author} {\bibfnamefont {D.~J.}\ \bibnamefont
			{Evans}}\ and\ \bibinfo {author} {\bibfnamefont {D.~J.}\ \bibnamefont
			{Searles}},\ }\href {\doibase 10.1080/00018730210155133} {\bibfield
		{journal} {\bibinfo  {journal} {Adv. Phys.}\ }\textbf {\bibinfo {volume}
			{51}},\ \bibinfo {pages} {1529} (\bibinfo {year} {2002})}\BibitemShut
	{NoStop}%
	\bibitem [{\citenamefont {Seifert}(2005)}]{Seifert.2005.PRL}%
	\BibitemOpen
	\bibfield  {author} {\bibinfo {author} {\bibfnamefont {U.}~\bibnamefont
			{Seifert}},\ }\href {\doibase 10.1103/PhysRevLett.95.040602} {\bibfield
		{journal} {\bibinfo  {journal} {Phys. Rev. Lett.}\ }\textbf {\bibinfo
			{volume} {95}},\ \bibinfo {pages} {040602} (\bibinfo {year}
		{2005})}\BibitemShut {NoStop}%
	\bibitem [{\citenamefont {Seifert}(2012)}]{Seifert.2012.RPP}%
	\BibitemOpen
	\bibfield  {author} {\bibinfo {author} {\bibfnamefont {U.}~\bibnamefont
			{Seifert}},\ }\href {http://stacks.iop.org/0034-4885/75/i=12/a=126001}
	{\bibfield  {journal} {\bibinfo  {journal} {Rep. Prog. Phys.}\ }\textbf
		{\bibinfo {volume} {75}},\ \bibinfo {pages} {126001} (\bibinfo {year}
		{2012})}\BibitemShut {NoStop}%
	\bibitem [{\citenamefont {Decker}(2015)}]{Yannick.2015.PA}%
	\BibitemOpen
	\bibfield  {author} {\bibinfo {author} {\bibfnamefont {Y.~D.}\ \bibnamefont
			{Decker}},\ }\href {\doibase 10.1016/j.physa.2015.01.073} {\bibfield
		{journal} {\bibinfo  {journal} {Physica A}\ }\textbf {\bibinfo {volume}
			{428}},\ \bibinfo {pages} {178 } (\bibinfo {year} {2015})}\BibitemShut
	{NoStop}%
	\bibitem [{\citenamefont {Fuchs}\ \emph {et~al.}(2016)\citenamefont {Fuchs},
		\citenamefont {Goldt},\ and\ \citenamefont {Seifert}}]{Jaco.2016.EPL}%
	\BibitemOpen
	\bibfield  {author} {\bibinfo {author} {\bibfnamefont {J.}~\bibnamefont
			{Fuchs}}, \bibinfo {author} {\bibfnamefont {S.}~\bibnamefont {Goldt}}, \ and\
		\bibinfo {author} {\bibfnamefont {U.}~\bibnamefont {Seifert}},\ }\href
	{http://stacks.iop.org/0295-5075/113/i=6/a=60009} {\bibfield  {journal}
		{\bibinfo  {journal} {EPL}\ }\textbf {\bibinfo {volume} {113}},\ \bibinfo
		{pages} {60009} (\bibinfo {year} {2016})}\BibitemShut {NoStop}%
	\bibitem [{\citenamefont {Gong}\ \emph {et~al.}(2016)\citenamefont {Gong},
		\citenamefont {Lan},\ and\ \citenamefont {Quan}}]{Gong.2016.PRL}%
	\BibitemOpen
	\bibfield  {author} {\bibinfo {author} {\bibfnamefont {Z.}~\bibnamefont
			{Gong}}, \bibinfo {author} {\bibfnamefont {Y.}~\bibnamefont {Lan}}, \ and\
		\bibinfo {author} {\bibfnamefont {H.~T.}\ \bibnamefont {Quan}},\ }\href
	{\doibase 10.1103/PhysRevLett.117.180603} {\bibfield  {journal} {\bibinfo
			{journal} {Phys. Rev. Lett.}\ }\textbf {\bibinfo {volume} {117}},\ \bibinfo
		{pages} {180603} (\bibinfo {year} {2016})}\BibitemShut {NoStop}%
	\bibitem [{\citenamefont {Goldt}\ and\ \citenamefont
		{Seifert}(2017)}]{Goldt.2017.PRL}%
	\BibitemOpen
	\bibfield  {author} {\bibinfo {author} {\bibfnamefont {S.}~\bibnamefont
			{Goldt}}\ and\ \bibinfo {author} {\bibfnamefont {U.}~\bibnamefont
			{Seifert}},\ }\href {\doibase 10.1103/PhysRevLett.118.010601} {\bibfield
		{journal} {\bibinfo  {journal} {Phys. Rev. Lett.}\ }\textbf {\bibinfo
			{volume} {118}},\ \bibinfo {pages} {010601} (\bibinfo {year}
		{2017})}\BibitemShut {NoStop}%
	\bibitem [{\citenamefont {Gallavotti}\ and\ \citenamefont
		{Cohen}(1995)}]{Gallavotti.1995.PRL}%
	\BibitemOpen
	\bibfield  {author} {\bibinfo {author} {\bibfnamefont {G.}~\bibnamefont
			{Gallavotti}}\ and\ \bibinfo {author} {\bibfnamefont {E.~G.~D.}\ \bibnamefont
			{Cohen}},\ }\href {\doibase 10.1103/PhysRevLett.74.2694} {\bibfield
		{journal} {\bibinfo  {journal} {Phys. Rev. Lett.}\ }\textbf {\bibinfo
			{volume} {74}},\ \bibinfo {pages} {2694} (\bibinfo {year}
		{1995})}\BibitemShut {NoStop}%
	\bibitem [{\citenamefont {Jarzynski}(1997)}]{Jarzynski.1997.PRL}%
	\BibitemOpen
	\bibfield  {author} {\bibinfo {author} {\bibfnamefont {C.}~\bibnamefont
			{Jarzynski}},\ }\href {\doibase 10.1103/PhysRevLett.78.2690} {\bibfield
		{journal} {\bibinfo  {journal} {Phys. Rev. Lett.}\ }\textbf {\bibinfo
			{volume} {78}},\ \bibinfo {pages} {2690} (\bibinfo {year}
		{1997})}\BibitemShut {NoStop}%
	\bibitem [{\citenamefont {Barato}\ and\ \citenamefont
		{Seifert}(2015{\natexlab{a}})}]{Barato.2015.PRL}%
	\BibitemOpen
	\bibfield  {author} {\bibinfo {author} {\bibfnamefont {A.~C.}\ \bibnamefont
			{Barato}}\ and\ \bibinfo {author} {\bibfnamefont {U.}~\bibnamefont
			{Seifert}},\ }\href {\doibase 10.1103/PhysRevLett.114.158101} {\bibfield
		{journal} {\bibinfo  {journal} {Phys. Rev. Lett.}\ }\textbf {\bibinfo
			{volume} {114}},\ \bibinfo {pages} {158101} (\bibinfo {year}
		{2015}{\natexlab{a}})}\BibitemShut {NoStop}%
	\bibitem [{\citenamefont {Gingrich}\ \emph {et~al.}(2016)\citenamefont
		{Gingrich}, \citenamefont {Horowitz}, \citenamefont {Perunov},\ and\
		\citenamefont {England}}]{Gingrich.2016.PRL}%
	\BibitemOpen
	\bibfield  {author} {\bibinfo {author} {\bibfnamefont {T.~R.}\ \bibnamefont
			{Gingrich}}, \bibinfo {author} {\bibfnamefont {J.~M.}\ \bibnamefont
			{Horowitz}}, \bibinfo {author} {\bibfnamefont {N.}~\bibnamefont {Perunov}}, \
		and\ \bibinfo {author} {\bibfnamefont {J.~L.}\ \bibnamefont {England}},\
	}\href {\doibase 10.1103/PhysRevLett.116.120601} {\bibfield  {journal}
		{\bibinfo  {journal} {Phys. Rev. Lett.}\ }\textbf {\bibinfo {volume} {116}},\
		\bibinfo {pages} {120601} (\bibinfo {year} {2016})}\BibitemShut {NoStop}%
	\bibitem [{\citenamefont {Pietzonka}\ \emph
		{et~al.}(2016{\natexlab{a}})\citenamefont {Pietzonka}, \citenamefont
		{Barato},\ and\ \citenamefont {Seifert}}]{Pietzonka.2016.PRE}%
	\BibitemOpen
	\bibfield  {author} {\bibinfo {author} {\bibfnamefont {P.}~\bibnamefont
			{Pietzonka}}, \bibinfo {author} {\bibfnamefont {A.~C.}\ \bibnamefont
			{Barato}}, \ and\ \bibinfo {author} {\bibfnamefont {U.}~\bibnamefont
			{Seifert}},\ }\href {\doibase 10.1103/PhysRevE.93.052145} {\bibfield
		{journal} {\bibinfo  {journal} {Phys. Rev. E}\ }\textbf {\bibinfo {volume}
			{93}},\ \bibinfo {pages} {052145} (\bibinfo {year}
		{2016}{\natexlab{a}})}\BibitemShut {NoStop}%
	\bibitem [{\citenamefont {Polettini}\ \emph {et~al.}(2016)\citenamefont
		{Polettini}, \citenamefont {Lazarescu},\ and\ \citenamefont
		{Esposito}}]{Polettini.2016.PRE}%
	\BibitemOpen
	\bibfield  {author} {\bibinfo {author} {\bibfnamefont {M.}~\bibnamefont
			{Polettini}}, \bibinfo {author} {\bibfnamefont {A.}~\bibnamefont
			{Lazarescu}}, \ and\ \bibinfo {author} {\bibfnamefont {M.}~\bibnamefont
			{Esposito}},\ }\href {\doibase 10.1103/PhysRevE.94.052104} {\bibfield
		{journal} {\bibinfo  {journal} {Phys. Rev. E}\ }\textbf {\bibinfo {volume}
			{94}},\ \bibinfo {pages} {052104} (\bibinfo {year} {2016})}\BibitemShut
	{NoStop}%
	\bibitem [{\citenamefont {Horowitz}\ and\ \citenamefont
		{Gingrich}(2017)}]{Horowitz.2017.PRE}%
	\BibitemOpen
	\bibfield  {author} {\bibinfo {author} {\bibfnamefont {J.~M.}\ \bibnamefont
			{Horowitz}}\ and\ \bibinfo {author} {\bibfnamefont {T.~R.}\ \bibnamefont
			{Gingrich}},\ }\href {\doibase 10.1103/PhysRevE.96.020103} {\bibfield
		{journal} {\bibinfo  {journal} {Phys. Rev. E}\ }\textbf {\bibinfo {volume}
			{96}},\ \bibinfo {pages} {020103(R)} (\bibinfo {year} {2017})}\BibitemShut
	{NoStop}%
	\bibitem [{\citenamefont {Dechant}\ and\ \citenamefont
		{Sasa}(2018)}]{Andreas.2018.JSM}%
	\BibitemOpen
	\bibfield  {author} {\bibinfo {author} {\bibfnamefont {A.}~\bibnamefont
			{Dechant}}\ and\ \bibinfo {author} {\bibfnamefont {S.-i.}\ \bibnamefont
			{Sasa}},\ }\href {http://stacks.iop.org/1742-5468/2018/i=6/a=063209}
	{\bibfield  {journal} {\bibinfo  {journal} {J. Stat. Mech.: Theory Exp.}\
		}\textbf {\bibinfo {volume} {2018}},\ \bibinfo {pages} {063209} (\bibinfo
		{year} {2018})}\BibitemShut {NoStop}%
	\bibitem [{\citenamefont {Mehta}\ and\ \citenamefont
		{Schwab}(2012)}]{Mehta.2012.PNAS}%
	\BibitemOpen
	\bibfield  {author} {\bibinfo {author} {\bibfnamefont {P.}~\bibnamefont
			{Mehta}}\ and\ \bibinfo {author} {\bibfnamefont {D.~J.}\ \bibnamefont
			{Schwab}},\ }\href {\doibase 10.1073/pnas.1207814109} {\bibfield  {journal}
		{\bibinfo  {journal} {Proc. Natl. Acad. Sci. U.S.A.}\ }\textbf {\bibinfo
			{volume} {109}},\ \bibinfo {pages} {17978} (\bibinfo {year}
		{2012})}\BibitemShut {NoStop}%
	\bibitem [{\citenamefont {Hasegawa}(2018)}]{Hasegawa.2018.PRE}%
	\BibitemOpen
	\bibfield  {author} {\bibinfo {author} {\bibfnamefont {Y.}~\bibnamefont
			{Hasegawa}},\ }\href {\doibase 10.1103/PhysRevE.98.032405} {\bibfield
		{journal} {\bibinfo  {journal} {Phys. Rev. E}\ }\textbf {\bibinfo {volume}
			{98}},\ \bibinfo {pages} {032405} (\bibinfo {year} {2018})}\BibitemShut
	{NoStop}%
	\bibitem [{\citenamefont {Barato}\ and\ \citenamefont
		{Seifert}(2015{\natexlab{b}})}]{Barato.2015.JPCB}%
	\BibitemOpen
	\bibfield  {author} {\bibinfo {author} {\bibfnamefont {A.~C.}\ \bibnamefont
			{Barato}}\ and\ \bibinfo {author} {\bibfnamefont {U.}~\bibnamefont
			{Seifert}},\ }\href {\doibase 10.1021/acs.jpcb.5b01918} {\bibfield  {journal}
		{\bibinfo  {journal} {J. Phys. Chem. B}\ }\textbf {\bibinfo {volume} {119}},\
		\bibinfo {pages} {6555} (\bibinfo {year} {2015}{\natexlab{b}})}\BibitemShut
	{NoStop}%
	\bibitem [{\citenamefont {Falasco}\ \emph {et~al.}(2016)\citenamefont
		{Falasco}, \citenamefont {Pfaller}, \citenamefont {Bregulla}, \citenamefont
		{Cichos},\ and\ \citenamefont {Kroy}}]{Falasco.2016.PRE}%
	\BibitemOpen
	\bibfield  {author} {\bibinfo {author} {\bibfnamefont {G.}~\bibnamefont
			{Falasco}}, \bibinfo {author} {\bibfnamefont {R.}~\bibnamefont {Pfaller}},
		\bibinfo {author} {\bibfnamefont {A.~P.}\ \bibnamefont {Bregulla}}, \bibinfo
		{author} {\bibfnamefont {F.}~\bibnamefont {Cichos}}, \ and\ \bibinfo {author}
		{\bibfnamefont {K.}~\bibnamefont {Kroy}},\ }\href {\doibase
		10.1103/PhysRevE.94.030602} {\bibfield  {journal} {\bibinfo  {journal} {Phys.
				Rev. E}\ }\textbf {\bibinfo {volume} {94}},\ \bibinfo {pages} {030602(R)}
		(\bibinfo {year} {2016})}\BibitemShut {NoStop}%
	\bibitem [{\citenamefont {Pietzonka}\ \emph
		{et~al.}(2016{\natexlab{b}})\citenamefont {Pietzonka}, \citenamefont
		{Barato},\ and\ \citenamefont {Seifert}}]{Patrick.2016.JSM}%
	\BibitemOpen
	\bibfield  {author} {\bibinfo {author} {\bibfnamefont {P.}~\bibnamefont
			{Pietzonka}}, \bibinfo {author} {\bibfnamefont {A.~C.}\ \bibnamefont
			{Barato}}, \ and\ \bibinfo {author} {\bibfnamefont {U.}~\bibnamefont
			{Seifert}},\ }\href {http://stacks.iop.org/1742-5468/2016/i=12/a=124004}
	{\bibfield  {journal} {\bibinfo  {journal} {J. Stat. Mech.: Theory Exp.}\
		}\textbf {\bibinfo {volume} {2016}},\ \bibinfo {pages} {124004} (\bibinfo
		{year} {2016}{\natexlab{b}})}\BibitemShut {NoStop}%
	\bibitem [{\citenamefont {Rotskoff}(2017)}]{Rotskoff.2017.PRE}%
	\BibitemOpen
	\bibfield  {author} {\bibinfo {author} {\bibfnamefont {G.~M.}\ \bibnamefont
			{Rotskoff}},\ }\href {\doibase 10.1103/PhysRevE.95.030101} {\bibfield
		{journal} {\bibinfo  {journal} {Phys. Rev. E}\ }\textbf {\bibinfo {volume}
			{95}},\ \bibinfo {pages} {030101(R)} (\bibinfo {year} {2017})}\BibitemShut
	{NoStop}%
	\bibitem [{\citenamefont {Garrahan}(2017)}]{Garrahan.2017.PRE}%
	\BibitemOpen
	\bibfield  {author} {\bibinfo {author} {\bibfnamefont {J.~P.}\ \bibnamefont
			{Garrahan}},\ }\href {\doibase 10.1103/PhysRevE.95.032134} {\bibfield
		{journal} {\bibinfo  {journal} {Phys. Rev. E}\ }\textbf {\bibinfo {volume}
			{95}},\ \bibinfo {pages} {032134} (\bibinfo {year} {2017})}\BibitemShut
	{NoStop}%
	\bibitem [{\citenamefont {Gingrich}\ and\ \citenamefont
		{Horowitz}(2017)}]{Gingrich.2017.PRL}%
	\BibitemOpen
	\bibfield  {author} {\bibinfo {author} {\bibfnamefont {T.~R.}\ \bibnamefont
			{Gingrich}}\ and\ \bibinfo {author} {\bibfnamefont {J.~M.}\ \bibnamefont
			{Horowitz}},\ }\href {\doibase 10.1103/PhysRevLett.119.170601} {\bibfield
		{journal} {\bibinfo  {journal} {Phys. Rev. Lett.}\ }\textbf {\bibinfo
			{volume} {119}},\ \bibinfo {pages} {170601} (\bibinfo {year}
		{2017})}\BibitemShut {NoStop}%
	\bibitem [{\citenamefont {Hyeon}\ and\ \citenamefont
		{Hwang}(2017)}]{Hyeon.2017.PRE}%
	\BibitemOpen
	\bibfield  {author} {\bibinfo {author} {\bibfnamefont {C.}~\bibnamefont
			{Hyeon}}\ and\ \bibinfo {author} {\bibfnamefont {W.}~\bibnamefont {Hwang}},\
	}\href {\doibase 10.1103/PhysRevE.96.012156} {\bibfield  {journal} {\bibinfo
			{journal} {Phys. Rev. E}\ }\textbf {\bibinfo {volume} {96}},\ \bibinfo
		{pages} {012156} (\bibinfo {year} {2017})}\BibitemShut {NoStop}%
	\bibitem [{\citenamefont {Proesmans}\ and\ \citenamefont {den
			Broeck}(2017)}]{Proesmans.2017.EPL}%
	\BibitemOpen
	\bibfield  {author} {\bibinfo {author} {\bibfnamefont {K.}~\bibnamefont
			{Proesmans}}\ and\ \bibinfo {author} {\bibfnamefont {C.~V.}\ \bibnamefont
			{den Broeck}},\ }\href {http://stacks.iop.org/0295-5075/119/i=2/a=20001}
	{\bibfield  {journal} {\bibinfo  {journal} {EPL}\ }\textbf {\bibinfo {volume}
			{119}},\ \bibinfo {pages} {20001} (\bibinfo {year} {2017})}\BibitemShut
	{NoStop}%
	\bibitem [{\citenamefont {Chiuchi\`u}\ and\ \citenamefont
		{Pigolotti}(2018)}]{Chiuchiu.2018.PRE}%
	\BibitemOpen
	\bibfield  {author} {\bibinfo {author} {\bibfnamefont {D.}~\bibnamefont
			{Chiuchi\`u}}\ and\ \bibinfo {author} {\bibfnamefont {S.}~\bibnamefont
			{Pigolotti}},\ }\href {\doibase 10.1103/PhysRevE.97.032109} {\bibfield
		{journal} {\bibinfo  {journal} {Phys. Rev. E}\ }\textbf {\bibinfo {volume}
			{97}},\ \bibinfo {pages} {032109} (\bibinfo {year} {2018})}\BibitemShut
	{NoStop}%
	\bibitem [{\citenamefont {Brandner}\ \emph {et~al.}(2018)\citenamefont
		{Brandner}, \citenamefont {Hanazato},\ and\ \citenamefont
		{Saito}}]{Brandner.2018.PRL}%
	\BibitemOpen
	\bibfield  {author} {\bibinfo {author} {\bibfnamefont {K.}~\bibnamefont
			{Brandner}}, \bibinfo {author} {\bibfnamefont {T.}~\bibnamefont {Hanazato}},
		\ and\ \bibinfo {author} {\bibfnamefont {K.}~\bibnamefont {Saito}},\ }\href
	{\doibase 10.1103/PhysRevLett.120.090601} {\bibfield  {journal} {\bibinfo
			{journal} {Phys. Rev. Lett.}\ }\textbf {\bibinfo {volume} {120}},\ \bibinfo
		{pages} {090601} (\bibinfo {year} {2018})}\BibitemShut {NoStop}%
	\bibitem [{\citenamefont {Hwang}\ and\ \citenamefont
		{Hyeon}(2018)}]{Hwang.2018.JPCL}%
	\BibitemOpen
	\bibfield  {author} {\bibinfo {author} {\bibfnamefont {W.}~\bibnamefont
			{Hwang}}\ and\ \bibinfo {author} {\bibfnamefont {C.}~\bibnamefont {Hyeon}},\
	}\href {\doibase 10.1021/acs.jpclett.7b03197} {\bibfield  {journal} {\bibinfo
			{journal} {J. Phys. Chem. Lett.}\ }\textbf {\bibinfo {volume} {9}},\
		\bibinfo {pages} {513} (\bibinfo {year} {2018})}\BibitemShut {NoStop}%
	\bibitem [{\citenamefont {Barato}\ \emph
		{et~al.}(2018{\natexlab{a}})\citenamefont {Barato}, \citenamefont {Chetrite},
		\citenamefont {Faggionato},\ and\ \citenamefont
		{Gabrielli}}]{Barato.2018.NJP}%
	\BibitemOpen
	\bibfield  {author} {\bibinfo {author} {\bibfnamefont {A.~C.}\ \bibnamefont
			{Barato}}, \bibinfo {author} {\bibfnamefont {R.}~\bibnamefont {Chetrite}},
		\bibinfo {author} {\bibfnamefont {A.}~\bibnamefont {Faggionato}}, \ and\
		\bibinfo {author} {\bibfnamefont {D.}~\bibnamefont {Gabrielli}},\ }\href
	{http://stacks.iop.org/1367-2630/20/i=10/a=103023} {\bibfield  {journal}
		{\bibinfo  {journal} {New J. Phys.}\ }\textbf {\bibinfo {volume} {20}},\
		\bibinfo {pages} {103023} (\bibinfo {year} {2018}{\natexlab{a}})}\BibitemShut
	{NoStop}%
	\bibitem [{\citenamefont {Barato}\ \emph
		{et~al.}(2018{\natexlab{b}})\citenamefont {Barato}, \citenamefont {Chetrite},
		\citenamefont {Faggionato},\ and\ \citenamefont
		{Gabrielli}}]{Barato.2018.arxiv}%
	\BibitemOpen
	\bibfield  {author} {\bibinfo {author} {\bibfnamefont {A.}~\bibnamefont
			{Barato}}, \bibinfo {author} {\bibfnamefont {R.}~\bibnamefont {Chetrite}},
		\bibinfo {author} {\bibfnamefont {A.}~\bibnamefont {Faggionato}}, \ and\
		\bibinfo {author} {\bibfnamefont {D.}~\bibnamefont {Gabrielli}},\ }\href
	{https://arxiv.org/abs/1810.11894} {\bibfield  {journal} {\bibinfo  {journal}
			{arXiv:1810.11894}\ } (\bibinfo {year} {2018}{\natexlab{b}})}\BibitemShut
	{NoStop}%
	\bibitem [{\citenamefont {Hasegawa}\ and\ \citenamefont
		{Van~Vu}(2019{\natexlab{a}})}]{Hasegawa.2019.PRE}%
	\BibitemOpen
	\bibfield  {author} {\bibinfo {author} {\bibfnamefont {Y.}~\bibnamefont
			{Hasegawa}}\ and\ \bibinfo {author} {\bibfnamefont {T.}~\bibnamefont
			{Van~Vu}},\ }\href {\doibase 10.1103/PhysRevE.99.062126} {\bibfield
		{journal} {\bibinfo  {journal} {Phys. Rev. E}\ }\textbf {\bibinfo {volume}
			{99}},\ \bibinfo {pages} {062126} (\bibinfo {year}
		{2019}{\natexlab{a}})}\BibitemShut {NoStop}%
	\bibitem [{\citenamefont {Koyuk}\ \emph {et~al.}(2019)\citenamefont {Koyuk},
		\citenamefont {Seifert},\ and\ \citenamefont {Pietzonka}}]{Koyuk.2019.JPA}%
	\BibitemOpen
	\bibfield  {author} {\bibinfo {author} {\bibfnamefont {T.}~\bibnamefont
			{Koyuk}}, \bibinfo {author} {\bibfnamefont {U.}~\bibnamefont {Seifert}}, \
		and\ \bibinfo {author} {\bibfnamefont {P.}~\bibnamefont {Pietzonka}},\ }\href
	{http://stacks.iop.org/1751-8121/52/i=2/a=02LT02} {\bibfield  {journal}
		{\bibinfo  {journal} {J. Phys. A: Math. Theor.}\ }\textbf {\bibinfo {volume}
			{52}},\ \bibinfo {pages} {02LT02} (\bibinfo {year} {2019})}\BibitemShut
	{NoStop}%
	\bibitem [{\citenamefont {Van~Vu}\ and\ \citenamefont
		{Hasegawa}(2019)}]{Vu.2019.arxiv}%
	\BibitemOpen
	\bibfield  {author} {\bibinfo {author} {\bibfnamefont {T.}~\bibnamefont
			{Van~Vu}}\ and\ \bibinfo {author} {\bibfnamefont {Y.}~\bibnamefont
			{Hasegawa}},\ }\href {https://arxiv.org/abs/1901.05715} {\bibfield  {journal}
		{\bibinfo  {journal} {arXiv:1901.05715}\ } (\bibinfo {year}
		{2019})}\BibitemShut {NoStop}%
	\bibitem [{\citenamefont {Bratsun}\ \emph {et~al.}(2005)\citenamefont
		{Bratsun}, \citenamefont {Volfson}, \citenamefont {Tsimring},\ and\
		\citenamefont {Hasty}}]{Bratsun.2005.PNAS}%
	\BibitemOpen
	\bibfield  {author} {\bibinfo {author} {\bibfnamefont {D.}~\bibnamefont
			{Bratsun}}, \bibinfo {author} {\bibfnamefont {D.}~\bibnamefont {Volfson}},
		\bibinfo {author} {\bibfnamefont {L.~S.}\ \bibnamefont {Tsimring}}, \ and\
		\bibinfo {author} {\bibfnamefont {J.}~\bibnamefont {Hasty}},\ }\href
	{\doibase 10.1073/pnas.0503858102} {\bibfield  {journal} {\bibinfo  {journal}
			{Proc. Natl. Acad. Sci. U.S.A.}\ }\textbf {\bibinfo {volume} {102}},\
		\bibinfo {pages} {14593} (\bibinfo {year} {2005})}\BibitemShut {NoStop}%
	\bibitem [{\citenamefont {Mather}\ \emph {et~al.}(2009)\citenamefont {Mather},
		\citenamefont {Bennett}, \citenamefont {Hasty},\ and\ \citenamefont
		{Tsimring}}]{Mather.2009.PRL}%
	\BibitemOpen
	\bibfield  {author} {\bibinfo {author} {\bibfnamefont {W.}~\bibnamefont
			{Mather}}, \bibinfo {author} {\bibfnamefont {M.~R.}\ \bibnamefont {Bennett}},
		\bibinfo {author} {\bibfnamefont {J.}~\bibnamefont {Hasty}}, \ and\ \bibinfo
		{author} {\bibfnamefont {L.~S.}\ \bibnamefont {Tsimring}},\ }\href {\doibase
		10.1103/PhysRevLett.102.068105} {\bibfield  {journal} {\bibinfo  {journal}
			{Phys. Rev. Lett.}\ }\textbf {\bibinfo {volume} {102}},\ \bibinfo {pages}
		{068105} (\bibinfo {year} {2009})}\BibitemShut {NoStop}%
	\bibitem [{\citenamefont {Nov{\'a}k}\ and\ \citenamefont
		{Tyson}(2008)}]{Novak.2008.NAT}%
	\BibitemOpen
	\bibfield  {author} {\bibinfo {author} {\bibfnamefont {B.}~\bibnamefont
			{Nov{\'a}k}}\ and\ \bibinfo {author} {\bibfnamefont {J.~J.}\ \bibnamefont
			{Tyson}},\ }\href {\doibase 10.1038/nrm2530} {\bibfield  {journal} {\bibinfo
			{journal} {Nat. Rev. Mol. Cell Biol.}\ }\textbf {\bibinfo {volume} {9}},\
		\bibinfo {pages} {981} (\bibinfo {year} {2008})}\BibitemShut {NoStop}%
	\bibitem [{\citenamefont {Kim}\ \emph {et~al.}(1999)\citenamefont {Kim},
		\citenamefont {Park},\ and\ \citenamefont {Pyo}}]{Kim.1999.PRL}%
	\BibitemOpen
	\bibfield  {author} {\bibinfo {author} {\bibfnamefont {S.}~\bibnamefont
			{Kim}}, \bibinfo {author} {\bibfnamefont {S.~H.}\ \bibnamefont {Park}}, \
		and\ \bibinfo {author} {\bibfnamefont {H.-B.}\ \bibnamefont {Pyo}},\ }\href
	{\doibase 10.1103/PhysRevLett.82.1620} {\bibfield  {journal} {\bibinfo
			{journal} {Phys. Rev. Lett.}\ }\textbf {\bibinfo {volume} {82}},\ \bibinfo
		{pages} {1620} (\bibinfo {year} {1999})}\BibitemShut {NoStop}%
	\bibitem [{\citenamefont {Masoller}(2003)}]{Masoller.2003.PRL}%
	\BibitemOpen
	\bibfield  {author} {\bibinfo {author} {\bibfnamefont {C.}~\bibnamefont
			{Masoller}},\ }\href {\doibase 10.1103/PhysRevLett.90.020601} {\bibfield
		{journal} {\bibinfo  {journal} {Phys. Rev. Lett.}\ }\textbf {\bibinfo
			{volume} {90}},\ \bibinfo {pages} {020601} (\bibinfo {year}
		{2003})}\BibitemShut {NoStop}%
	\bibitem [{\citenamefont {Lichtner}\ \emph {et~al.}(2012)\citenamefont
		{Lichtner}, \citenamefont {Pototsky},\ and\ \citenamefont
		{Klapp}}]{Lichtner.2012.PRE}%
	\BibitemOpen
	\bibfield  {author} {\bibinfo {author} {\bibfnamefont {K.}~\bibnamefont
			{Lichtner}}, \bibinfo {author} {\bibfnamefont {A.}~\bibnamefont {Pototsky}},
		\ and\ \bibinfo {author} {\bibfnamefont {S.~H.~L.}\ \bibnamefont {Klapp}},\
	}\href {\doibase 10.1103/PhysRevE.86.051405} {\bibfield  {journal} {\bibinfo
			{journal} {Phys. Rev. E}\ }\textbf {\bibinfo {volume} {86}},\ \bibinfo
		{pages} {051405} (\bibinfo {year} {2012})}\BibitemShut {NoStop}%
	\bibitem [{\citenamefont {Loos}\ and\ \citenamefont
		{Klapp}(2019)}]{Loos.2019.SR}%
	\BibitemOpen
	\bibfield  {author} {\bibinfo {author} {\bibfnamefont {S.~A.~M.}\
			\bibnamefont {Loos}}\ and\ \bibinfo {author} {\bibfnamefont {S.~H.~L.}\
			\bibnamefont {Klapp}},\ }\href {\doibase 10.1038/s41598-019-39320-0}
	{\bibfield  {journal} {\bibinfo  {journal} {Sci. Rep.}\ }\textbf {\bibinfo
			{volume} {9}},\ \bibinfo {pages} {2491} (\bibinfo {year} {2019})}\BibitemShut
	{NoStop}%
	\bibitem [{\citenamefont {Rosinberg}\ \emph {et~al.}(2015)\citenamefont
		{Rosinberg}, \citenamefont {Munakata},\ and\ \citenamefont
		{Tarjus}}]{Rosinberg.2015.PRE}%
	\BibitemOpen
	\bibfield  {author} {\bibinfo {author} {\bibfnamefont {M.~L.}\ \bibnamefont
			{Rosinberg}}, \bibinfo {author} {\bibfnamefont {T.}~\bibnamefont {Munakata}},
		\ and\ \bibinfo {author} {\bibfnamefont {G.}~\bibnamefont {Tarjus}},\ }\href
	{\doibase 10.1103/PhysRevE.91.042114} {\bibfield  {journal} {\bibinfo
			{journal} {Phys. Rev. E}\ }\textbf {\bibinfo {volume} {91}},\ \bibinfo
		{pages} {042114} (\bibinfo {year} {2015})}\BibitemShut {NoStop}%
	\bibitem [{\citenamefont {Rosinberg}\ \emph {et~al.}(2017)\citenamefont
		{Rosinberg}, \citenamefont {Tarjus},\ and\ \citenamefont
		{Munakata}}]{Rosinberg.2017.PRE}%
	\BibitemOpen
	\bibfield  {author} {\bibinfo {author} {\bibfnamefont {M.~L.}\ \bibnamefont
			{Rosinberg}}, \bibinfo {author} {\bibfnamefont {G.}~\bibnamefont {Tarjus}}, \
		and\ \bibinfo {author} {\bibfnamefont {T.}~\bibnamefont {Munakata}},\ }\href
	{\doibase 10.1103/PhysRevE.95.022123} {\bibfield  {journal} {\bibinfo
			{journal} {Phys. Rev. E}\ }\textbf {\bibinfo {volume} {95}},\ \bibinfo
		{pages} {022123} (\bibinfo {year} {2017})}\BibitemShut {NoStop}%
	\bibitem [{\citenamefont {Frank}\ \emph {et~al.}(2003)\citenamefont {Frank},
		\citenamefont {Beek},\ and\ \citenamefont {Friedrich}}]{Frank.2003.PRE}%
	\BibitemOpen
	\bibfield  {author} {\bibinfo {author} {\bibfnamefont {T.~D.}\ \bibnamefont
			{Frank}}, \bibinfo {author} {\bibfnamefont {P.~J.}\ \bibnamefont {Beek}}, \
		and\ \bibinfo {author} {\bibfnamefont {R.}~\bibnamefont {Friedrich}},\ }\href
	{\doibase 10.1103/PhysRevE.68.021912} {\bibfield  {journal} {\bibinfo
			{journal} {Phys. Rev. E}\ }\textbf {\bibinfo {volume} {68}},\ \bibinfo
		{pages} {021912} (\bibinfo {year} {2003})}\BibitemShut {NoStop}%
	\bibitem [{\citenamefont {Frank}(2005)}]{Frank.2005.PRE}%
	\BibitemOpen
	\bibfield  {author} {\bibinfo {author} {\bibfnamefont {T.~D.}\ \bibnamefont
			{Frank}},\ }\href {\doibase 10.1103/PhysRevE.72.011112} {\bibfield  {journal}
		{\bibinfo  {journal} {Phys. Rev. E}\ }\textbf {\bibinfo {volume} {72}},\
		\bibinfo {pages} {011112} (\bibinfo {year} {2005})}\BibitemShut {NoStop}%
	\bibitem [{\citenamefont {Gomez-Marin}\ \emph {et~al.}(2008)\citenamefont
		{Gomez-Marin}, \citenamefont {Parrondo},\ and\ \citenamefont {Van~den
			Broeck}}]{GomezMarin.2008.PRE}%
	\BibitemOpen
	\bibfield  {author} {\bibinfo {author} {\bibfnamefont {A.}~\bibnamefont
			{Gomez-Marin}}, \bibinfo {author} {\bibfnamefont {J.~M.~R.}\ \bibnamefont
			{Parrondo}}, \ and\ \bibinfo {author} {\bibfnamefont {C.}~\bibnamefont
			{Van~den Broeck}},\ }\href {\doibase 10.1103/PhysRevE.78.011107} {\bibfield
		{journal} {\bibinfo  {journal} {Phys. Rev. E}\ }\textbf {\bibinfo {volume}
			{78}},\ \bibinfo {pages} {011107} (\bibinfo {year} {2008})}\BibitemShut
	{NoStop}%
	\bibitem [{\citenamefont {Rold\'an}\ and\ \citenamefont
		{Parrondo}(2012)}]{Roldan.2012.PRE}%
	\BibitemOpen
	\bibfield  {author} {\bibinfo {author} {\bibfnamefont {E.}~\bibnamefont
			{Rold\'an}}\ and\ \bibinfo {author} {\bibfnamefont {J.~M.~R.}\ \bibnamefont
			{Parrondo}},\ }\href {\doibase 10.1103/PhysRevE.85.031129} {\bibfield
		{journal} {\bibinfo  {journal} {Phys. Rev. E}\ }\textbf {\bibinfo {volume}
			{85}},\ \bibinfo {pages} {031129} (\bibinfo {year} {2012})}\BibitemShut
	{NoStop}%
	\bibitem [{\citenamefont {Diana}\ and\ \citenamefont
		{Esposito}(2014)}]{Diana.2014.JSM}%
	\BibitemOpen
	\bibfield  {author} {\bibinfo {author} {\bibfnamefont {G.}~\bibnamefont
			{Diana}}\ and\ \bibinfo {author} {\bibfnamefont {M.}~\bibnamefont
			{Esposito}},\ }\href {\doibase 10.1088/1742-5468/2014/04/p04010} {\bibfield
		{journal} {\bibinfo  {journal} {J. Stat. Mech.: Theory Exp.}\ }\textbf
		{\bibinfo {volume} {2014}},\ \bibinfo {pages} {P04010} (\bibinfo {year}
		{2014})}\BibitemShut {NoStop}%
	\bibitem [{\citenamefont {Gaspard}(2004)}]{Gaspard.2004.JSP}%
	\BibitemOpen
	\bibfield  {author} {\bibinfo {author} {\bibfnamefont {P.}~\bibnamefont
			{Gaspard}},\ }\href {\doibase 10.1007/s10955-004-3455-1} {\bibfield
		{journal} {\bibinfo  {journal} {J. Stat. Phys.}\ }\textbf {\bibinfo {volume}
			{117}},\ \bibinfo {pages} {599} (\bibinfo {year} {2004})}\BibitemShut
	{NoStop}%
	\bibitem [{\citenamefont {Trepagnier}\ \emph {et~al.}(2004)\citenamefont
		{Trepagnier}, \citenamefont {Jarzynski}, \citenamefont {Ritort},
		\citenamefont {Crooks}, \citenamefont {Bustamante},\ and\ \citenamefont
		{Liphardt}}]{Trepagnier.2004.PNAS}%
	\BibitemOpen
	\bibfield  {author} {\bibinfo {author} {\bibfnamefont {E.~H.}\ \bibnamefont
			{Trepagnier}}, \bibinfo {author} {\bibfnamefont {C.}~\bibnamefont
			{Jarzynski}}, \bibinfo {author} {\bibfnamefont {F.}~\bibnamefont {Ritort}},
		\bibinfo {author} {\bibfnamefont {G.~E.}\ \bibnamefont {Crooks}}, \bibinfo
		{author} {\bibfnamefont {C.~J.}\ \bibnamefont {Bustamante}}, \ and\ \bibinfo
		{author} {\bibfnamefont {J.}~\bibnamefont {Liphardt}},\ }\href
	{https://doi.org/10.1073/pnas.0406405101} {\bibfield  {journal} {\bibinfo
			{journal} {Proc. Natl. Acad. Sci. U.S.A.}\ }\textbf {\bibinfo {volume}
			{101}},\ \bibinfo {pages} {15038} (\bibinfo {year} {2004})}\BibitemShut
	{NoStop}%
	\bibitem [{\citenamefont {Collin}\ \emph {et~al.}(2005)\citenamefont {Collin},
		\citenamefont {Ritort}, \citenamefont {Jarzynski}, \citenamefont {Smith},
		\citenamefont {Tinoco~Jr},\ and\ \citenamefont
		{Bustamante}}]{Collin.2005.Nature}%
	\BibitemOpen
	\bibfield  {author} {\bibinfo {author} {\bibfnamefont {D.}~\bibnamefont
			{Collin}}, \bibinfo {author} {\bibfnamefont {F.}~\bibnamefont {Ritort}},
		\bibinfo {author} {\bibfnamefont {C.}~\bibnamefont {Jarzynski}}, \bibinfo
		{author} {\bibfnamefont {S.~B.}\ \bibnamefont {Smith}}, \bibinfo {author}
		{\bibfnamefont {I.}~\bibnamefont {Tinoco~Jr}}, \ and\ \bibinfo {author}
		{\bibfnamefont {C.}~\bibnamefont {Bustamante}},\ }\href
	{https://doi.org/10.1038/nature04061} {\bibfield  {journal} {\bibinfo
			{journal} {Nature}\ }\textbf {\bibinfo {volume} {437}},\ \bibinfo {pages}
		{231} (\bibinfo {year} {2005})}\BibitemShut {NoStop}%
	\bibitem [{\citenamefont {Schuler}\ \emph {et~al.}(2005)\citenamefont
		{Schuler}, \citenamefont {Speck}, \citenamefont {Tietz}, \citenamefont
		{Wrachtrup},\ and\ \citenamefont {Seifert}}]{Schuler.2005.PRL}%
	\BibitemOpen
	\bibfield  {author} {\bibinfo {author} {\bibfnamefont {S.}~\bibnamefont
			{Schuler}}, \bibinfo {author} {\bibfnamefont {T.}~\bibnamefont {Speck}},
		\bibinfo {author} {\bibfnamefont {C.}~\bibnamefont {Tietz}}, \bibinfo
		{author} {\bibfnamefont {J.}~\bibnamefont {Wrachtrup}}, \ and\ \bibinfo
		{author} {\bibfnamefont {U.}~\bibnamefont {Seifert}},\ }\href
	{https://doi.org/10.1103/PhysRevLett.94.180602} {\bibfield  {journal}
		{\bibinfo  {journal} {Phys. Rev. Lett.}\ }\textbf {\bibinfo {volume} {94}},\
		\bibinfo {pages} {180602} (\bibinfo {year} {2005})}\BibitemShut {NoStop}%
	\bibitem [{\citenamefont {Tietz}\ \emph {et~al.}(2006)\citenamefont {Tietz},
		\citenamefont {Schuler}, \citenamefont {Speck}, \citenamefont {Seifert},\
		and\ \citenamefont {Wrachtrup}}]{Tietz.2006.PRL}%
	\BibitemOpen
	\bibfield  {author} {\bibinfo {author} {\bibfnamefont {C.}~\bibnamefont
			{Tietz}}, \bibinfo {author} {\bibfnamefont {S.}~\bibnamefont {Schuler}},
		\bibinfo {author} {\bibfnamefont {T.}~\bibnamefont {Speck}}, \bibinfo
		{author} {\bibfnamefont {U.}~\bibnamefont {Seifert}}, \ and\ \bibinfo
		{author} {\bibfnamefont {J.}~\bibnamefont {Wrachtrup}},\ }\href
	{https://doi.org/10.1103/PhysRevLett.97.050602} {\bibfield  {journal}
		{\bibinfo  {journal} {Phys. Rev. Lett.}\ }\textbf {\bibinfo {volume} {97}},\
		\bibinfo {pages} {050602} (\bibinfo {year} {2006})}\BibitemShut {NoStop}%
	\bibitem [{\citenamefont {Andrieux}\ \emph {et~al.}(2007)\citenamefont
		{Andrieux}, \citenamefont {Gaspard}, \citenamefont {Ciliberto}, \citenamefont
		{Garnier}, \citenamefont {Joubaud},\ and\ \citenamefont
		{Petrosyan}}]{Andrieux.2007.PRL}%
	\BibitemOpen
	\bibfield  {author} {\bibinfo {author} {\bibfnamefont {D.}~\bibnamefont
			{Andrieux}}, \bibinfo {author} {\bibfnamefont {P.}~\bibnamefont {Gaspard}},
		\bibinfo {author} {\bibfnamefont {S.}~\bibnamefont {Ciliberto}}, \bibinfo
		{author} {\bibfnamefont {N.}~\bibnamefont {Garnier}}, \bibinfo {author}
		{\bibfnamefont {S.}~\bibnamefont {Joubaud}}, \ and\ \bibinfo {author}
		{\bibfnamefont {A.}~\bibnamefont {Petrosyan}},\ }\href
	{https://doi.org/10.1103/PhysRevLett.98.150601} {\bibfield  {journal}
		{\bibinfo  {journal} {Phys. Rev. Lett.}\ }\textbf {\bibinfo {volume} {98}},\
		\bibinfo {pages} {150601} (\bibinfo {year} {2007})}\BibitemShut {NoStop}%
	\bibitem [{\citenamefont {Aron}\ \emph {et~al.}(2010)\citenamefont {Aron},
		\citenamefont {Biroli},\ and\ \citenamefont {Cugliandolo}}]{Aron.2010.JSM}%
	\BibitemOpen
	\bibfield  {author} {\bibinfo {author} {\bibfnamefont {C.}~\bibnamefont
			{Aron}}, \bibinfo {author} {\bibfnamefont {G.}~\bibnamefont {Biroli}}, \ and\
		\bibinfo {author} {\bibfnamefont {L.~F.}\ \bibnamefont {Cugliandolo}},\
	}\href {\doibase 10.1088/1742-5468/2010/11/p11018} {\bibfield  {journal}
		{\bibinfo  {journal} {J. Stat. Mech.: Theory Exp.}\ }\textbf {\bibinfo
			{volume} {2010}},\ \bibinfo {pages} {P11018} (\bibinfo {year}
		{2010})}\BibitemShut {NoStop}%
	\bibitem [{\citenamefont {Hasegawa}\ and\ \citenamefont
		{Van~Vu}(2019{\natexlab{b}})}]{Hasegawa.2019.arxiv}%
	\BibitemOpen
	\bibfield  {author} {\bibinfo {author} {\bibfnamefont {Y.}~\bibnamefont
			{Hasegawa}}\ and\ \bibinfo {author} {\bibfnamefont {T.}~\bibnamefont
			{Van~Vu}},\ }\href {https://arxiv.org/abs/1902.06376} {\bibfield  {journal}
		{\bibinfo  {journal} {arXiv:1902.06376}\ } (\bibinfo {year}
		{2019}{\natexlab{b}})}\BibitemShut {NoStop}%
	\bibitem [{\citenamefont {Merhav}\ and\ \citenamefont
		{Kafri}(2010)}]{Merhav.2010.JSM}%
	\BibitemOpen
	\bibfield  {author} {\bibinfo {author} {\bibfnamefont {N.}~\bibnamefont
			{Merhav}}\ and\ \bibinfo {author} {\bibfnamefont {Y.}~\bibnamefont {Kafri}},\
	}\href {\doibase 10.1088/1742-5468/2010/12/p12022} {\bibfield  {journal}
		{\bibinfo  {journal} {J. Stat. Mech.: Theory Exp.}\ }\textbf {\bibinfo
			{volume} {2010}},\ \bibinfo {pages} {P12022} (\bibinfo {year}
		{2010})}\BibitemShut {NoStop}%
	\bibitem [{\citenamefont {Shiraishi}(2017)}]{Shiraishi.2017.arxiv}%
	\BibitemOpen
	\bibfield  {author} {\bibinfo {author} {\bibfnamefont {N.}~\bibnamefont
			{Shiraishi}},\ }\href {https://arxiv.org/abs/1706.00892} {\bibfield
		{journal} {\bibinfo  {journal} {arXiv:1706.00892}\ } (\bibinfo {year}
		{2017})}\BibitemShut {NoStop}%
	\bibitem [{\citenamefont {K{\"u}chler}\ and\ \citenamefont
		{Mensch}(1992)}]{Kuchler.1992.SSR}%
	\BibitemOpen
	\bibfield  {author} {\bibinfo {author} {\bibfnamefont {U.}~\bibnamefont
			{K{\"u}chler}}\ and\ \bibinfo {author} {\bibfnamefont {B.}~\bibnamefont
			{Mensch}},\ }\href {\doibase 10.1080/17442509208833780} {\bibfield  {journal}
		{\bibinfo  {journal} {Stochastics and Stochastic Rep.}\ }\textbf {\bibinfo
			{volume} {40}},\ \bibinfo {pages} {23} (\bibinfo {year} {1992})}\BibitemShut
	{NoStop}%
	\bibitem [{\citenamefont {Sekimoto}(2010)}]{Sekimoto.2010}%
	\BibitemOpen
	\bibfield  {author} {\bibinfo {author} {\bibfnamefont {K.}~\bibnamefont
			{Sekimoto}},\ }\href@noop {} {\emph {\bibinfo {title} {Stochastic
				Energetics}}},\ Vol.\ \bibinfo {volume} {799}\ (\bibinfo  {publisher}
	{Springer},\ \bibinfo {year} {2010})\BibitemShut {NoStop}%
	\bibitem [{\citenamefont {van Zon}\ and\ \citenamefont
		{Cohen}(2003)}]{Zon.2003.PRE}%
	\BibitemOpen
	\bibfield  {author} {\bibinfo {author} {\bibfnamefont {R.}~\bibnamefont {van
				Zon}}\ and\ \bibinfo {author} {\bibfnamefont {E.~G.~D.}\ \bibnamefont
			{Cohen}},\ }\href {\doibase 10.1103/PhysRevE.67.046102} {\bibfield  {journal}
		{\bibinfo  {journal} {Phys. Rev. E}\ }\textbf {\bibinfo {volume} {67}},\
		\bibinfo {pages} {046102} (\bibinfo {year} {2003})}\BibitemShut {NoStop}%
	\bibitem [{\citenamefont {Mai}\ and\ \citenamefont
		{Dhar}(2007)}]{Mai.2007.PRE}%
	\BibitemOpen
	\bibfield  {author} {\bibinfo {author} {\bibfnamefont {T.}~\bibnamefont
			{Mai}}\ and\ \bibinfo {author} {\bibfnamefont {A.}~\bibnamefont {Dhar}},\
	}\href {\doibase 10.1103/PhysRevE.75.061101} {\bibfield  {journal} {\bibinfo
			{journal} {Phys. Rev. E}\ }\textbf {\bibinfo {volume} {75}},\ \bibinfo
		{pages} {061101} (\bibinfo {year} {2007})}\BibitemShut {NoStop}%
	\bibitem [{\citenamefont {Paredes-Altuve}\ \emph {et~al.}(2016)\citenamefont
		{Paredes-Altuve}, \citenamefont {Medina},\ and\ \citenamefont
		{Colmenares}}]{Paredes.2016.PRE}%
	\BibitemOpen
	\bibfield  {author} {\bibinfo {author} {\bibfnamefont {O.}~\bibnamefont
			{Paredes-Altuve}}, \bibinfo {author} {\bibfnamefont {E.}~\bibnamefont
			{Medina}}, \ and\ \bibinfo {author} {\bibfnamefont {P.~J.}\ \bibnamefont
			{Colmenares}},\ }\href {\doibase 10.1103/PhysRevE.94.062111} {\bibfield
		{journal} {\bibinfo  {journal} {Phys. Rev. E}\ }\textbf {\bibinfo {volume}
			{94}},\ \bibinfo {pages} {062111} (\bibinfo {year} {2016})}\BibitemShut
	{NoStop}%
	\bibitem [{\citenamefont {Ghosh}\ and\ \citenamefont
		{Chaudhury}(2017)}]{Ghosh.2017.PA}%
	\BibitemOpen
	\bibfield  {author} {\bibinfo {author} {\bibfnamefont {B.}~\bibnamefont
			{Ghosh}}\ and\ \bibinfo {author} {\bibfnamefont {S.}~\bibnamefont
			{Chaudhury}},\ }\href {\doibase https://doi.org/10.1016/j.physa.2016.09.001}
	{\bibfield  {journal} {\bibinfo  {journal} {Physica A}\ }\textbf {\bibinfo
			{volume} {466}},\ \bibinfo {pages} {133 } (\bibinfo {year}
		{2017})}\BibitemShut {NoStop}%
	\bibitem [{\citenamefont {Ohkuma}\ and\ \citenamefont
		{Ohta}(2007)}]{Ohkuma.2007.JSM}%
	\BibitemOpen
	\bibfield  {author} {\bibinfo {author} {\bibfnamefont {T.}~\bibnamefont
			{Ohkuma}}\ and\ \bibinfo {author} {\bibfnamefont {T.}~\bibnamefont {Ohta}},\
	}\href {\doibase 10.1088/1742-5468/2007/10/p10010} {\bibfield  {journal}
		{\bibinfo  {journal} {J. Stat. Mech.: Theory Exp.}\ }\textbf {\bibinfo
			{volume} {2007}},\ \bibinfo {pages} {P10010} (\bibinfo {year}
		{2007})}\BibitemShut {NoStop}%
	\bibitem [{\citenamefont {Rosinberg}\ and\ \citenamefont
		{Horowitz}(2016)}]{Rosinberg.2016.EPL}%
	\BibitemOpen
	\bibfield  {author} {\bibinfo {author} {\bibfnamefont {M.~L.}\ \bibnamefont
			{Rosinberg}}\ and\ \bibinfo {author} {\bibfnamefont {J.~M.}\ \bibnamefont
			{Horowitz}},\ }\href {\doibase 10.1209/0295-5075/116/10007} {\bibfield
		{journal} {\bibinfo  {journal} {EPL}\ }\textbf {\bibinfo {volume} {116}},\
		\bibinfo {pages} {10007} (\bibinfo {year} {2016})}\BibitemShut {NoStop}%
	\bibitem [{\citenamefont {Risken}(1989)}]{Risken.1989}%
	\BibitemOpen
	\bibfield  {author} {\bibinfo {author} {\bibfnamefont {H.}~\bibnamefont
			{Risken}},\ }\href@noop {} {\emph {\bibinfo {title} {The Fokker-Planck
				Equation: Methods of Solution and Applications}}},\ \bibinfo {edition} {2nd}\
	ed.\ (\bibinfo  {publisher} {Springer},\ \bibinfo {year} {1989})\BibitemShut
	{NoStop}%
\end{thebibliography}
\end{document}